\documentclass[
amsmath,amssymb,aps, pra, reprint, superscriptaddress,
floatfix, nofootinbib ]{revtex4-1}

\usepackage[all,arc]{xy}
\usepackage{enumerate}
\usepackage{mathtools,amsmath,amssymb,amsfonts,amsthm,xfrac,enumitem,pstricks,pst-node,float,xy,mathrsfs,accents,hhline,graphicx,color}
\usepackage[normalem]{ulem}
\xyoption{all}

\usepackage{graphicx,subcaption}
\usepackage{dcolumn}
\usepackage{bm}

\def\>{\rangle}
\def\<{\langle}

\def\t{\dagger}

\def\->{\rightarrow}
\def\=>{\implies}
\def\6{\partial}

\newcommand{\Rom}[1]{\expandafter\@slowromancap\romannumeral #1@}

\graphicspath{{./Figures/}}

\begin{document}

\preprint{APS/123-QED}

\title{Locally optimal measurement-based quantum feedback with application to multi-qubit entanglement generation}

\author{Song Zhang$^\dagger$}
\affiliation{Berkeley Center for Quantum Information and Computation, Berkeley, California 94720, USA}
\affiliation{Department of Physics, University of California, Berkeley, California 94720, USA}
\author{Leigh S. Martin$^\dagger$}
\affiliation{Berkeley Center for Quantum Information and Computation, Berkeley, California 94720, USA}
\affiliation{Department of Physics, University of California, Berkeley, California 94720, USA}
\author{K. Birgitta Whaley}
\email{corresponding author: whaley@berkeley.edu}
\affiliation{Berkeley Center for Quantum Information and Computation, Berkeley, California 94720, USA}
\affiliation{Department of Chemistry, University of California, Berkeley, California 94720, USA}

\date{\today}

\begin{abstract}
We present a general approach to measurement-based quantum feedback that employs proportional and quantum state-based (PaQS) feedback components to obtain locally optimal protocols. To demonstrate the power of the method, we first show that it reproduces many known feedback protocols, and then apply it to generation of multipartite entanglement with an emphasis on remote entanglement, which requires spatially local feedback Hamiltonians.  The symmetry of both measurement and feedback operators is found to be essential for construction of effective protocols. We show that under perfect measurement efficiency, entangled states can be reached with fidelity approaching unity under non-Markovian feedback control protocols, while Markovian protocols resulting from optimizing the feedback unitaries on ensemble averaged states still yield fidelities above 94\%. Application of the PaQS approach to generation of N-qubit W, general Dicke and GHZ states shows that such entangled states can be efficiently generated with high fidelity, for up to $N=100$ in some cases.
\end{abstract}

\maketitle

Entanglement is a crucial resource for quantum information science, with applications in secure cryptography~\cite{Ekert1991}, long-range quantum state transfer~\cite{Kimble2008},
quantum computation ~\cite{Wendin2016}, 
quantum-enhanced sensors~\cite{Wineland1994,Kitagawa1993,Hosten2016}
and quantum simulation~\cite{Bloch2012}. Many applications, particularly
large-scale quantum information processing, will require modular quantum devices that are able to talk to each other~\cite{Devoret2013superconducting,Benjamin:2013}.
However, the objects that we would like to entangle often have little or no direct interaction to enable entanglement generation, either because they are separated by significant distances, or because of intrinsically weak physical interactions. The former is the case in most quantum networking applications, as well as in nitrogen vacancy centers, for which large inter-qubit spacing is necessary to maintain coherence properties \cite{hanson2008coherent}.  The latter is the case for dilute neutral atoms, which have received considerable interest for generation of spin-squeezed states involving large numbers of entangled atoms~\cite{Stockton2004,Cox2016}.

When no direct interaction is available, joint measurement on all parties offers a practical method to project the total system into a desired entangled state. This has been demonstrated by interfering spontaneously emitted optical photons emitted from (artificial) atomic systems~\cite{Chou2005,Moehring2007,Hofmann2012,Bernien2013}, and also by  performing cavity-QED-based joint measurements on superconducting qubits~\cite{Roch:2014ey}	
or cold neutral atoms~\cite{takano2009spin}.

However, often the stochastic nature of a quantum measurement prevents such an approach from succeeding with unit probability target ~\cite{note:measurem_probab}. 
In order to improve the probability of reaching an entangled state, or to obtain it deterministically, one can then add control via feedback unitary operations. Several experimental and theoretical works have shown that this approach can yield deterministic remote entanglement generation between two qubits~\cite{Martin2015,Martin2017a,Riste:2013um}.
Multi-partite entanglement generation has been experimentally demonstrated in Ref. \cite{Cox2016},
with a feedback strategy for deterministically preparing an ensemble of atoms near the maximally spin-squeezed subspace.  
Generation of multi-spin states has also been discussed in the theoretical control literature in the context of state stabilization~\cite{vanHandel2005,Mirrahimi2007,Tsumura2008}. 
Refs. \cite{Thomsen2002}, \cite{Stockton2004}, and \cite{Wei2015} have proposed state-based \textit{i.e.,} non-Markovian protocols for deterministic preparation of Dicke states. In the conclusion of Ref. \cite{Stockton2004}, the authors emphasize both the experimental difficulty of implementing non-Markovian protocols and the importance of constructive methods to derive feedback control laws, but leave these issues as open problems for future research.

In the current work, we address both of these issues with a general method for constructing locally optimal measurement-based feedback protocols. 
Measurement-based feedback protocols rely on the fact that realistic measurements acquire only partial information over a finite time interval. Such protocols can in principle be applied to any system in which the measurement signal is collected with high efficiency, and have proven particularly useful in recent years for superconducting qubits ~\cite{Vijay2011,Vijay:2012ua,Riste:2013um,Murch:2013ur,Roch:2014ey,Hacohen-Gourgy2016}. In these systems, the finite time interval is taken to be small enough that the measurements may be regarded as continuous and feedback consists of applying additional control unitary operations conditional on the outcome of the continuous measurement.

Our approach to design locally optimal protocols is based on the introduction of time-dependent feedback unitaries composed of two independently varying components, one of which depends linearly on the measurement outcome with a state-dependent coefficient, while the other depends only on the quantum state. We term this approach ``Proportional and Quantum State" (PaQS) feedback, to emphasize the increased flexibility offered by these two independent feedback components.
We also show how this construction can be modified to guarantee Markovianity, meaning that each feedback operation only depends on the immediately proceeding measurement outcome, which greatly simplifies experimental implementation. We demonstrate this formalism by first using it to reproduce several well-established measurement-based feedback protocols, and then giving a systematic treatment of entanglement generation in three-qubit systems and beyond, emphasizing the crucial role that symmetry plays in the protocols in order to achieve high fidelities. We consider  generation of both the $N$-qubit GHZ states and the full range of $N$-qubit symmetric Dicke states,  from W states to the half-filled states, which are maximally spin-squeezed. Our protocols are characterized by a high degree of symmetry and are derived using general analytic techniques that allow us study up to 100 qubits. These advantages also confer remarkably simple experimental implementations of the feedback controller.

The remainder of the paper is organized as follows. In section \ref{sec:cont_meas}, we first summarize the evolution of a general quantum state under continuous measurement and feedback as described by the stochastic master equations, and then present our 
PaQS formalism for efficiently computing locally optimal protocols in the context of a general measurement-based feedback system.

In sections \ref{sec:W} and \ref{sec:GHZ} we apply the PaQS formalism to the task of generating entangled many-qubit states, considering specifically Dicke states and GHZ states. 
In section \ref{subsec:multioperatorGHZ} we then show that the PaQS formulation of section \ref{sec:cont_meas} can be extended to systems undergoing multiple distinct measurements and commuting feedback Hamiltonians, and use this to obtain a second protocol for GHZ states.  A third protocol for GHZ states that is based on optimizing an entanglement measure rather than a target state fidelity is presented in appendix \ref{app:non-markovian}.
Section~\ref{sec:deterministic} shows how under certain conditions the feedback can be chosen to eliminate the randomness introduced by the ``quantum noise" inherent to the measurement process. We introduce a strict condition for deterministic evolution, given fixed measurement and feedback operators and show that this provides a simple tool for finding solutions to proportional feedback master equations in general.
Section \ref{sec:con} provides a summary of the results and an outlook for future work. Key calculational details and supplementary materials are presented in the Appendices.

\begin{figure}[t]
\includegraphics[width=0.5\textwidth]{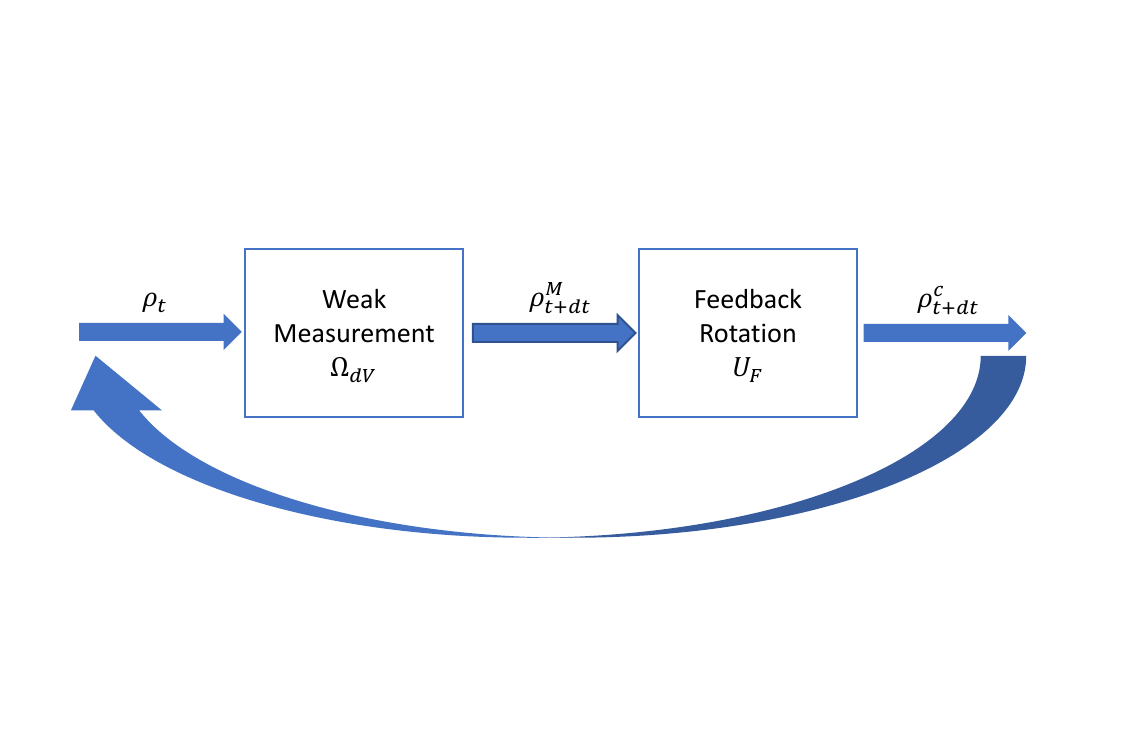}
\caption{A diagrammatic picture illustrating what happens during a measurement-feedback cycle over time $dt$.}
\label{fig:feedback_cycle}
\end{figure}

\section{\label{sec:cont_meas}Continuous measurement with locally optimal feedback} 
\subsection{Continuous measurement and feedback}

We consider a joint measurement that is realized by an indirect simultaneous measurement on multiple qubits. Such measurements are routinely made for superconducting qubits in the dispersive regime, using homodyne detection of cavity transmission ~\cite{Gambetta2008a, Murch:2013ur,Roch:2014ey,Riste:2013um}.  The readout voltage of the signal is given by 
\begin{equation}\label{eqn:measurementoutcome}
dV=\<X\>(t)dt+\frac{dW}{\sqrt{8\eta k}}
\end{equation}
where $X$ is the measurement operator which will be specified in detail below, $\<\cdot\>$ denotes the trace average of this, $dW$ is a Wiener increment satisfying $dW(t)dW(t') = \delta(t-t') dt$ that represents the quantum uncertainty in the homodyne detection (represented by white Gaussian noise in the continuous limit~\cite{Wiseman:2009vw,Oksendal2003}), $\eta$ is the measurement efficiency, and $k$ is the measurement strength.

The evolution of the quantum state conditioned on this measurement signal is given by a stochastic master equation (SME), given by 
\begin{equation}\label{eqn:measurement}
d\rho=-\frac{i}{\hbar}[H_S,\rho]dt + 2k \mathcal{D}[X]\rho(t)dt + \sqrt{2k\eta}\mathcal{H}[X]\rho dW.
\end{equation}
Here $H_S$ is the system Hamiltonian which is not relevant to this analysis and will henceforth be set to $0$, $\mathcal{D}[X]\rho=X\rho X^\dagger-\frac{1}{2}(X^\dagger X\rho+\rho X^\dagger X)$, and $\mathcal{H}[X]\rho=X\rho +\rho X^\dagger -\<X+X^\dagger\>\rho$. The second, deterministic, term of order $dt$ describes the dephasing effect of the measurement and is just the usual dissipator term in the Markovian master equation. It describes the unconditioned dynamics of an open system coupled to an external bath, \textit{i.e.}, the dynamics after averaging over all possible measurement records. The third, stochastic, term of order $dW$ continuously updates $\rho$ based on knowledge gained from the measurement.

As the qubits are assumed to be too remote to allow direct interactions, we construct a unitary feedback operator in the form of local operations on each of the $N$ qubits,
\begin{align} \label{eqn:feedback}
U_F  &=\bigotimes_{i=1}^N U_i. 
\end{align}
In the first instance we shall employ
\begin{align} \label{eqn:feedback_singlequbit}
U_i  &= e^{-i \theta_i H_i /\hbar},
\end{align}
with $H_i = \frac{1}{2}\hat{n}\cdot\vec{\sigma}$, where $(\theta_i,\hat{n}_i)$ are rotation angle parameters for the $i$th qubit. The single qubit unitaries are then simply rotation operators on a single qubit, $U_i=e^{-i\frac{\theta}{2}\hat{n}\cdot\vec{\sigma}/\hbar}$, with 
$\vec{\sigma}=(\sigma_x, \sigma_y, \sigma_z)$ denoting the usual Pauli operators. These rotation angles will be determined based on the measurement result. For simplicity, we assume that these rotations are realized instantaneously. 

After adding the feedback, the complete evolution of the qubits over a single infinitesimal cycle of measurement and feedback (see Fig. \ref{fig:feedback_cycle}) is described by the following:
\begin{equation}\label{eqn: output_state}
\rho_{t+dt}^c=U_F\rho_{t+dt}^M U_F^\dagger=U_F(\rho_t+d\rho)U_F^\dagger
\end{equation}
where $d\rho$ is taken from Eq.~\eqref{eqn:measurement}, the superscript $M$ indicates it is the conditioned state resulting from the measurement, and the superscript $c$ indicates that this state is obtained after the feedback control. 

\subsection{General locally optimal control protocol with PaQS}
\label{subsec:locallyoptimal}
Now we are ready to state the locally optimal control problem in complete generality and to find the solution. Suppose we have a state $\rho_t$ at time $t$, and that we ultimately wish to reach some target state $|\psi_T\rangle$. In this work we shall focus primarily on the fidelity with respect to the target state as the cost function, whose optimization determines the parameter of the feedback operator. (Appendix \ref{app:non-markovian} presents optimization under an alternative, non-linear cost function given by an entanglement measure.) 
For the entangled states of interest in later sections, i.e., Dicke and GHZ states, we shall take all values of $\theta_i$ to be equal, to ensure that the permutation symmetry of the state is maintained by the feedback operations (sections \ref{sec:W} and \ref{sec:GHZ}) although we shall relax this assumption when considering feedback with multiple, simultaneous measurements (section \ref{subsec:multioperatorGHZ}). Our feedback unitary is then
\begin{align} \label{eqn:feedback_rotation}
U_F(\theta) = e^{-i \theta H_F/\hbar}.
\end{align}
For the time being, we place no constraints on $H_F$, although for practical settings it will often be useful to take this to be separable.
The fidelity after a feedback operation is given by
\begin{align}
\mathcal{F}_{t+dt}(\theta) = \langle \psi_T|\rho^c(\theta)|\psi_T\rangle, \\ \nonumber
\rho^c(\theta) \equiv U(\theta)(\rho+d\rho)U^\dagger(\theta).
\end{align}
For convenience, we shall set $\hbar = 1$ from now on. Our local optimality condition is given by
\begin{align}\label{eqn:optimal_condition}
\mathcal{G} & \equiv \frac{\partial\mathcal{F}_{t+dt}(\theta)}{\partial \theta}  = \<\psi_T|\Big[U_F'(\theta)(\rho_t+d\rho)U_F^\dagger(\theta) + h.c.\Big]|\psi_T\>\\ \nonumber
& = -i\<\psi_T|[H_F,\rho^c]|\psi_T\> = 0
\end{align}
with $U'_F$ the derivative of $U_F$.

Using the fact that $d\rho$ is infinitesimal, we can derive an analytical expression for the optimal rotation angle $\theta$, which we denote $\theta^*$.
Assuming that $\rho_t$ is already optimized from the previous time step, then typically $\theta^*$ will also be $\mathcal{O}(d\rho)$ (we deal with possible exceptions below). We can therefore parameterize the rotation
angle as 
\begin{equation} \label{eqn:theta_parameterization}
\theta^* = A_1(t)dW + A_2(t)dt.
\end{equation}
Expanding $U_F$ to second order in $dW$ and making use of Ito's lemma~\cite{Oksendal2003} yields
\begin{equation}
U\equiv I-iA_1 H_F dW-(i A_2 H_F +\frac{1}{2}A_1^2 H_F^2)dt.
\end{equation}

Expanding Eq.~\eqref{eqn: output_state} and using the measurement stochastic master equation Eq.~\eqref{eqn:measurement} for $d\rho$, together with this second order expansion, yields the master equation 
\begin{align}\label{eqn:WM}
d\rho^c &= \mathcal{D}[Y]\rho dt + \mathcal{H}[Y]\rho dW -i (A_1 dW + A_2 dt)[H_F,\rho] \nonumber \\
& + A_1^2 \mathcal{D}[H_F]\rho dt - i {\sqrt{\eta}} A_1[H_F,Y\rho + \rho Y^\dagger]dt,
\end{align}
where we have defined $Y = \sqrt{2k}X$ to suppress $k$ in the result.
Employing $\rho^c$ in Eq.~\eqref{eqn:optimal_condition} leads to the following explicit form for $\mathcal{G}$:
\begin{widetext}
\begin{align} \label{eq:optimal_condition_2}
\mathcal{G} = & -i\langle \psi_T|[H_F, \rho_t]|\psi_T\rangle - i\langle \psi_T|[H_F, \mathcal{D}[Y]\rho_t dt + {\sqrt{\eta}}\mathcal{H}[Y]\rho_t dW]|\psi_T\rangle -A_1 \langle \psi_T|[H_F, [H_F, \rho_t]]|\psi_T\rangle dW \\ \nonumber 
& -i A_1^2\langle \psi_T|[H_F, \mathcal{D}[H_F]\rho_t]|\psi_T\rangle dt - \langle \psi_T|[H_F, [H_F, \sqrt{\eta} A_1\mathcal{H}[Y]\rho_t+A_2\rho_t]]|\psi_T\rangle dt, \nonumber
\end{align}
\end{widetext}
We now solve Eq.~\eqref{eq:optimal_condition_2} order by order in $dW$. Despite the 
large number of terms, it is nevertheless possible to solve for $A_1$ and $A_2$ in complete generality. 
The assumption that the optimal rotation was applied at the immediately preceding time step implies that
$\partial\mathcal{F}_t(\theta)/\partial \theta|_{\theta=0} = -i \langle \psi_T|[H_F,\rho_t]|\psi_T\rangle = 0$, so that the first term in Eq.~\eqref{eq:optimal_condition_2} may be dropped. Terms proportional to $dW$ yield a linear equation in $A_1$, which is easily solved. Once $A_1$ is known, terms proportional to $dW^2=dt$ are gathered to yield another linear equation, this time for $A_2$. The final result in full form is
\begin{widetext}
\begin{align}\label{eqn:optimal_coefficient}
&A_1 = \frac{-i\langle \psi_T|[H_F, \sqrt{\eta}(Y\rho_t+\rho_t Y^\dagger)]|\psi_T\rangle}{\langle\psi_T|[H_F, [H_F, \rho_t]]|\psi_T\rangle} \\ \nonumber
&A_2 = \frac{-\langle \psi_T|[H_F,i\mathcal{D}[Y]\rho_t+ \sqrt{\eta}A_1[H_F,Y\rho_t+\rho_t Y^\dagger]+i A_1^2 \mathcal{D}[H_F]\rho_t]|\psi_T\rangle}{\langle \psi_T|[H_F, [H_F, \rho_t]]|\psi_T\rangle}.
\end{align}
\end{widetext}
Using Eq.~\eqref{eqn:measurementoutcome} and \eqref{eqn:theta_parameterization}, the locally optimal feedback rotation can also be written as
\begin{align}\label{eqn:optimal_angle}
\theta^* = \sqrt{8\eta k}A_1 dV + (A_2 - 2A_1 \<Y\>) dt,
\end{align}
from which we see that $A_1$ can be identified with a proportional feedback term,
while the second term -  dependent on $A_1, A_2$, and the state at time $t$ - can be identified with an additional time dependent effective Hamiltonian drive. These two terms motivate the designation of ``PaQS".

Note that as $A_2$ only appears in Eq.~\eqref{eqn:WM} to first order, $\partial(d\mathcal{F})/\partial A_2$ is not a function of $A_2$, so then the naive method of computing the locally optimal protocol by maximizing $d\mathcal{F} \equiv \langle \psi_T|d\rho|\psi_T\rangle$ with respect to $A_2$ would not yield Eq.~\eqref{eqn:optimal_coefficient}.

So far, we have assumed that the optimal angle is infinitesimal. However, Eq.~\eqref{eqn:optimal_coefficient} only guarantees that the solution $\theta^*$ is a local extremum and does not guarantee that it is necessarily a maximum. A sufficient condition for $\theta^*$ to be a local maximum is that the second derivative of the fidelity function evaluated at $\theta^*$ be negative:
\begin{equation}\label{eqn:2DTest}
\frac{\partial^2\mathcal{F}_{t+dt}(\theta)}{\partial \theta^2} \bigg|_{\theta=\theta^*}= -\<\psi_T|[H_F,[H_F,\rho^c]]|\psi_T\>\bigg|_{\theta=\theta^*}<0.
\end{equation}
Failure of this test, \textit{i.e.}, when the second derivative is positive, suggests the presence of a local minimum from the infinitesimal solution. Then we will need a large (\textit{i.e.}, non-infinitesimal) rotation, which we compute by maximizing the fidelity over the entire angular range. 

It should be noted that $A_1$ and $A_2$ can in principle become singular. However as $\rho^c_{t-dt}|_{\theta^*} = \rho_t$, the denominator diverges only when the second derivative test failed at the previous time step (compare Eq.~\eqref{eqn:2DTest} to the denominator of Eq. \eqref{eqn:optimal_coefficient}). Thus this divergence is typically prevented by the global search described above. A special case is $[H_F, \rho_t]=0$, in which case feedback has no effect on the state, so that we may simply set $\theta^*=0$.

\begin{table*}
\begin{center}
\begin{tabular}{|c|c|c|c|}
\hline
\textbf{Feedback} & \textbf{Measurement} & \textbf{Feedback} & \textbf{Target state} \\
\textbf{protocol} & \textbf{operator} ($X$) & \textbf{Hamiltonian} ($H_F$) & ($|\psi_T\>$) \\
\hline
Adaptive phase & $\sigma$ & $\sigma_z$ (Heisenberg & $(|0\rangle + i|1\rangle)/\sqrt{2}$ \\
measurement \cite{Wis-1995}$^*$ &  &  picture) &\\
\hline
Rapid qubit & $\sigma_z$ & $\sigma_y$ & $(|g\rangle + |e\rangle)/\sqrt{2}$ \\
purification \cite{Jacobs:2003hc}$^*$ & & &\\
\hline
Half-parity Bell & $\sigma_{z,1}+\sigma_{z,2}$ & $\sigma_{y,1} + \sigma_{y,2}$ & $(|eg\> + |ge\>)/\sqrt{2}$ \\
state preparation \cite{Martin2015}$^*$ & & &\\
\hline
Full-parity Bell state & $\sigma_{z,1}\sigma_{z,2}$ & $\sigma_{x,1}$ & $(|gg\> + i|eg\>$ \\
 preparation \cite{Hill2008, Martin2017a}$^*$& & & $+i|eg\>+|ee\>)/2$ \\
\hline
N-qubit Dicke states & $\sum_i \sigma_{z,i}$ & $\sum_i \sigma_{y,i}$ & N-qubit Dicke state\\
 \cite{Thomsen2002,Stockton2004,Wei2015}$^\dagger$ & & & with n excitations \\
\hline
N-qubit GHZ & $\sum_{i \ne j}\sigma_{z,i}\sigma_{z,j}$ & $\sum_i \sigma_{x,i}$ & $N$-qubit GHZ state \\
states$^\dagger$ &  & \\ \hline
N-qubit GHZ & $\sigma_{z,i}-\sigma_{z,j}$ & $\sigma_{y,i}-\sigma_{y,j}$ & $N$-qubit GHZ state \\
states$^*$$^\dagger$ & for all $i,j$ & for all $i,j$ & \\ \hline
Hong-Ou-Mandel & $i(\sigma_1+\sigma_2)$,  & $\sigma_{x,1}+\sigma_{x,2}$ & $(|gg\>+|ee\>)/\sqrt{2}$ \\ 
Bell state \cite{martin2019single}$^*$ & $\sigma_1-\sigma_2$ & $\sigma_{y,1}-\sigma_{y,2}$ & \\ \hline
\end{tabular}
\end{center}
\caption{Summary of some measurement-based feedback protocols that may be derived from Eq.~\eqref{eqn:optimal_coefficient} or its generalization to multiple measurement and feedback operators in section \ref{sec:GHZ}. $^\dagger$ denotes this work. Protocols in which perfect noise cancellation (cancellation of $dW$ terms, see discussion in section \ref{sec:deterministic}) can occur are marked with an asterisk. Note that for $N$-qubit GHZ states, noise cancellation occurs only for the multiple measurement operator protocol with $N=2$ and $N=3$.}
\label{tab:PaQSTable}
\end{table*}

To simulate this form of feedback in practice, we assume that the controller chooses the rotation angle $\theta^*$ that ensures a global maximum of $\mathcal{F}_{t=0}$ at the initial time step. During evolution of the state, the above protocol typically continues to pick $\theta^*$ as the global maximum of $\mathcal{F}_t$ and thus maintains the system on a locally (time-)optimal trajectory. However even if Eq. \eqref{eqn:2DTest} remains negative, it is possible that the nearest local maximum of $\mathcal{F}_t(\theta)$ can fail to be the global maximum. The only way to catch such instances is to occasionally undertake a brute force maximization of $\mathcal{F}$ over the full range of $\theta$ and to thereby check whether the local maximum identified by Eq. \eqref{eqn:optimal_coefficient} is also a global maximum. In practice, such global maximization procedures are often unnecessary. Table \ref{tab:PaQSTable} shows many combinations of measurement operators $X$ and feedback Hamiltonians $H_F$ that allow Eqs. \eqref{eqn:optimal_coefficient} and \eqref{eqn:2DTest} to reproduce the indicated feedback protocols established in the control literature. Global searches are only required when indicated by Eq. \eqref{eqn:2DTest}. We shall see examples of this in the following section.

\begin{figure}[t]
\includegraphics[width=0.5\textwidth]{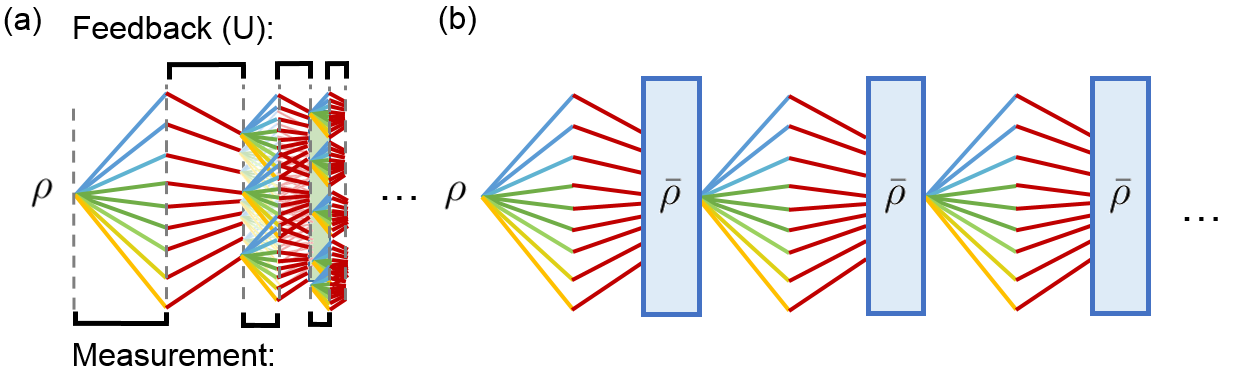}
\caption{(a) Schematic of a non-Markovian feedback protocol. The number of possible states, and hence the number of potentially distinct feedback operations needed in response scale exponentially in time. We refer to this as the trajectory ensemble approach (TEA). (b) Schematic of the average state locally optimal (ASLO) feedback protocol, which applies controls that are determined using only the average time-evolved state and the most recent measurement outcome  \cite{Martin2015}. This is the behavior of a ``forgetful" or Markovian controller.}
\label{fig:ASLO}
\end{figure}

The functions $A_1$ and $A_2$ are dependent on both the initial and target states, as well as on the state at time $t$. Dependence on the current state implies implicit dependence on the full measurement record, yielding a potentially non-Markovian feedback protocol in general. In experimental situations, non-Markovian protocols are significantly more difficult to implement than Markovian protocols. The feedback controller must either calculate $\rho(t)$ in real time, which is both challenging and time-consuming, or else perform an exponential amount of precomputation to determine the optimal action for all possible measurement records. This is illustrated schematically in Fig. \ref{fig:ASLO}a. However we can also use the feedback master equation to simulate a memoryless controller, which ensures that all realizations have the same feedback, regardless of the individual measurement records. This is the average state locally optimal (ASLO) protocol of Ref.~\cite{Martin2015}. The basic concept is illustrated in Fig.~\ref{fig:ASLO}b. 

In an ASLO protocol, instead of using the controlled state $\rho^c_{t+dt}$ of Eq. \eqref{eqn: output_state} to determine the feedback operation for the next time step, we use an unconditioned state that is obtained by averaging over the entire measurement record at time $t$. We express the evolution of the conditioned state as
\begin{equation} \label{eqn: conditioned_state}
\rho_{t+dt}^M=\frac{\Omega_{dV}\rho_t\Omega_{dV}^\dagger}{Tr(\Omega_{dV}\rho_t\Omega_{dV}^\dagger)},
\end{equation}
where $\Omega_{dV}$ is the POVM for continuous measurement with $I=\int dV\ \Omega_{dV}^\dagger\Omega_{dV}$:\footnote{Technically, this POVM is applicable only to an ideal measurement. However, inefficient measurement can be treated as limited access to multiple ideal measurement channels. The analysis below can be thereby readily extended to inefficient measurements.} 
\begin{equation} \label{eq:POVM}
\Omega_{dV}=\left(\frac{4k}{\pi dt}\right)^{\frac{1}{4}} \exp \left[-2k dt \left(\frac{dV}{dt}-X\right)^2\right ].
\end{equation}
Eqs.~\eqref{eqn: conditioned_state} and \eqref{eq:POVM} are equivalent to the stochastic master equation of Eq.~\ref{eqn:measurement} with $H_S=0$. Averaging over the measurement record at time $t$ then gives the unconditioned state
\begin{align}\label{eqn:ASLO_output_state}
\bar{\rho}_{t+dt} &\equiv  \int_{-\infty}^\infty d(dV_t) P(dV_t) \rho^c_{t+dt},
\end{align}
where $P(dV_t) = \textrm{Tr}(\Omega_{dV}\bar{\rho}_t\Omega_{dV}^\dagger)$ and $\rho^c_{t+dt}$ is implicitly dependent on $dV_t$ (Eq.~\eqref{eqn: output_state}). Inspection of Eq.~\eqref{eqn: output_state} shows that the averaged state $\bar{\rho}_{t+dt}$ is equal to 
\begin{equation} \label{eqn:ASLOandUF}
\bar{\rho}_{t+dt} = \int_{-\infty}^\infty d(dV)\    U_F\Omega_{dV}\bar{\rho}_t\Omega_{dV}^\dagger U_F^\dagger.
\end{equation}
At the end of the next time step $t+2dt$, the feedback controller then computes $\theta^*$ based on this unconditioned averaged state $\bar{\rho}_{t+dt}$ instead of based on $\rho_{t+dt}$. Iterating this procedure results in replacing $\rho$ by $\bar{\rho}$ in Eq.~\eqref{eqn:optimal_coefficient}. 
The state $\bar{\rho}_t$ along the evolution is then understood as an averaged state. It has  deterministic dynamics, as can be seen by averaging over, (\textit{i.e.}, dropping, since $\langle dW \rangle =0$) the terms proportional to $dW$ in the second order expansion Eq.~\eqref{eqn:WM}. This  
allows efficient simulation of an arbitrary ASLO feedback protocol by a single trajectory. The feedback protocol is now essentially Markovian, since the dependence on the previous measurement history has been removed by the averaging.   

The averaging procedure in Eq.~\eqref{eqn:ASLO_output_state} is a mathematical step that corresponds exactly to the averaging over trajectories with different measurement records and hence over quantum noise histories that is done in an experiment.  The feedback unitary characterized by $A_1(t)$ and $A_2(t)$ for the averaged state can then be applied to any individual realization of the state at each instant. This provides significant advantages for experimental implementation, since the functions $A_1(t)$ and $A_2(t)$ can be pre-calculated efficiently with the same procedure as for the TEA approach above but using a single calculated trajectory for $\bar{\rho}(t)$, and the resulting feedback operation applied to each experimental trajectory without the need for real-time state estimation.

Finally, we note that although the general optimization formalism above has been illustrated using fidelity with respect to a desired target state as the cost function, it can straight-forwardly be applied to other cost functions that are linear in $\rho$, such as the expectation value of an operator. In addition, one can also go beyond linear protocols to consider alternative cost functions based on entanglement measures. 

In the next two sections we apply the PaQS protocol to $N$-qubit  Dicke states (section~\ref{sec:W}) and to $N$-qubit GHZ states (section~\ref{sec:GHZ}). These two canonical examples of entangled $N$-qubit states are simple generalizations of the 2-qubit Bell states. For simplicity, we shall employ perfect measurement efficiency, $\eta = 1$, in all simulations throughout the rest of this paper, although the theory described in this section is valid also for inefficient measurements with $\eta < 1$. 

\section{Generation of Dicke states}\label{sec:W}
The Dicke states are defined as
\begin{equation}
\label{eqn:Dicke}
|N,k\> = \frac{1}{\sqrt{\binom{N}{k}}}\Sigma_{P\in S_N} P(|0\>^{\otimes (N-k)}\otimes |1\>^{\otimes k})
\end{equation}
where $P$ is an operator belonging to the permutation group $S_N$ on $N$ qubits. When $k=1$, we have the well-known W state. When $k = \frac{N}{2}$ ($N$ even or $k = \frac{N+1}{2}$ when $N$ is odd), we have the half-filled Dicke state.
$|N,n\>$ is a uniform superposition over all states with the same number of excitations, and is also known as a spin squeezed state. The Dicke states have been proposed for applications in a wide range of sensing protocols, including very long baseline interferometry \cite{gottesman2012longer} and Heisenberg-limited measurement sensitivity \cite{gross2012spin}.  

We shall first consider the canonical W state with $N=3$, before addressing the generation of general Dicke states for arbitrary $N$ values.
\subsection{Generation of the $N=3$ W state}
The $N=3$ W state, given by 
\begin{equation}
|W\>=\frac{1}{\sqrt{3}}(|001\>+|010\>+|100\>),
\end{equation}
represents a generic type of weak three qubit entanglement that is  characterized by zero three-way entanglement but maximal retention of bipartite entanglement on loss of a qubit~\cite{Dur2002}. For simplicity, in the rest of this section we shall refer to the $N=3$ state just as the W state. To prepare this state, we first choose a measurement observable 
\begin{equation}\label{eqn:w_observable}
X_{W}=\sigma_{z,1} + \sigma_{z,2} + \sigma_{z,3},
\end{equation}
where $\sigma_{z,i}$ is the Pauli operator along the $\hat{z}_i$ axis for qubit $i$. 
Note that this measurement operator has the same permutation symmetry as the W state. It is also a linear combination of single-body observables, which is the easiest form to implement on spatially separated qubits, in particular, on remote qubits, such as superconducting qubits in different microwave cavities.

The W state is an eigenstate of $X_{W}$, with eigenvalue $+1$. This implies that for any  initial state $\rho_0$, the long time limit of evolution under continuous measurement of $X_{W}$ in the absence of feedback will result in projection onto the W state with probability $p_W = \<W|\rho_0|W\>$. 

To increase this success probability, we apply the fidelity-optimized protocol of the previous section, choosing the target state as $|\psi_T\> = |W\>$. To choose a proper feedback rotation operator,
we generalize the locally optimal two-qubit feedback unitary used to generate the 2-qubit W state $\left[|10\rangle + |01\rangle\right]/\sqrt{2}$ in Ref.~\cite{Martin2015}, which was shown to provide a globally optimal protocol for both maximal fidelity and concurrence~\cite{Martin2017a}.  As in that work, we restrict the evolution of each qubit to lie in the $xz$ plane ($\phi_i =0, i=\{1,2,3\}$), with local rotations around the $y$ axis having equal angles for each of the three qubits, \textit{i.e.}, 
$\theta_1=\theta_2=\theta_3=\theta$. 
The latter condition is consistent with the fact that both the target $W$ state and the observable $X_W$ are symmetric with respect to any permutation of the three qubits. For the W state, the rotation operator in Eq.~\eqref{eqn:feedback_rotation} then takes the following form
\begin{equation}
 U_F^W(\theta) = e^{-i\frac{\theta}{2}(\sigma_{y,1}+\sigma_{y,1}+\sigma_{y,3})},
\label{eqn:w_unitary_F}
\end{equation}
where $\sigma_{y,i}$ are the Pauli operators along the $\hat{y}_i$ axis for qubit $i$. 

Now the permutation symmetry of both measurement $X_W$ and feedback  $U_F^W(\theta)$ operators induces a symmetric subspace 
$\mathcal{H}_W^S$ that is spanned by
\begin{align} \label{eqn:HF_subspace} \nonumber
|\phi_W^1\>&=|000\>\\ \nonumber
|\phi_W^2\>&=\frac{1}{\sqrt{3}}(|001\> + |010\>+ |100\>)\\ \nonumber
|\phi_W^3\>&=\frac{1}{\sqrt{3}}(|011\>+|101\>+|110\>)\\ 
|\phi_W^4\>&=|111\>
\end{align}
Thus, if our initial state is also symmetric to all permutations of the qubits, then the state will only evolve in the subspace $\mathcal{H}_W^S$ under the action of $X_W$ and $U_F^W(\theta)$. To see this, consider an arbitrary operator $P$ from the permutation group $S_3$. This clearly commutes with both $X_W$ and $U_F^W(\theta)$ in Eqs. \eqref{eqn:w_observable} and \eqref{eqn:w_unitary_F}. Then we can use Eq.~\eqref{eqn: output_state} to obtain
\begin{equation}\label{eqn:evolveP}
\begin{aligned}
P\rho_{t+dt}^c P &=  PU_F^W \rho_{t+dt}^M  U_F^{W\dagger}P = U_F^W P \rho_{t+dt}^M P  U_F^{W\dagger} \\ 
& =U_F^W (P\rho_{t}P + Pd\rho P) U_F^{W\dagger} \\
& = U_F^W (\rho_{t} + d\rho) U_F^{W\dagger}\\
&=\rho_{t+dt}^c,
\end{aligned}
\end{equation}
where to arrive at the final equality, we have used the assumption that $\rho_t$ is in the subspace $\mathcal{H}_W^S$ and the fact that $X_W$ commutes with $P$. Provided that the initial state $\rho(0)$ is in $\mathcal{H}_W^S$, this assumption will hold at any time $t$.
Now we wish to control the evolution so that the quantum state evolves from the initial state to the W state, which is $|\phi_W^2\>$. Noting that the basis vectors in Eq.~\eqref{eqn:HF_subspace} constitute non-degenerate eigenspaces of the measurement operator $X_W$ in the symmetric subspace, we can then achieve the desired evolution by implementing a feedback protocol in which the control angle $\theta$ is determined by a cost function that specifically targets the state $|\phi_W^2\>$ within $\mathcal{H}_W^S$.  The fidelity  $\mathcal{F}_W =  \<\phi_W^2|\rho_t|\psi_W^2\rangle$ is a suitable cost function for this.
\begin{figure}[t]
\includegraphics[width=0.5\textwidth]{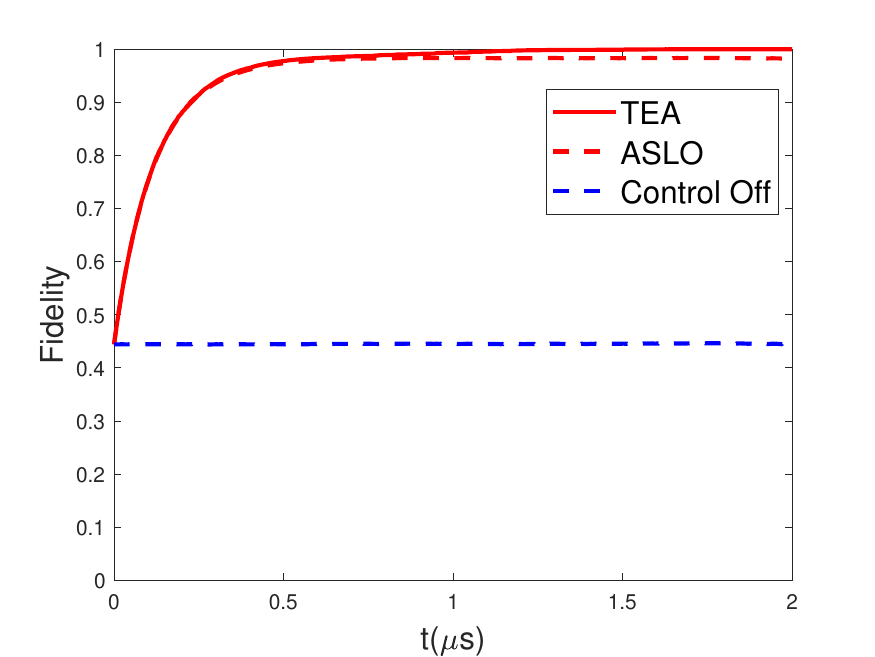}
\caption{
Fidelity of controlled state with respect to the W state, $\mathcal{F}_W$ starting from the initial product state $|000\>$. The solid red line shows results calculated with the trajectory ensemble approach (TEA) and the dashed red line shows results calculated with the ASLO approach. The dashed blue line shows the results obtained with measurement alone, \textit{i.e.}, in absence of feedback, after an initial rotation to the state of maximum fidelity with the W state that is achievable by $U_F^W(\theta)$. 
Over $1000$ trajectories were averaged for TEA, and over $10,000$ trajectories for the ASLO and zero feedback calculations. The measurement strength is set to $k=1$ MHz, corresponding to realistic values for superconducting qubits~\cite{Martin2015,Murch:2013ur}, and the time step is set  at $\Delta t \ll \frac{1}{k}$.}
\label{fig:W_fidelity}
\end{figure}

\begin{figure}[ht]
\includegraphics[width=0.5\textwidth]{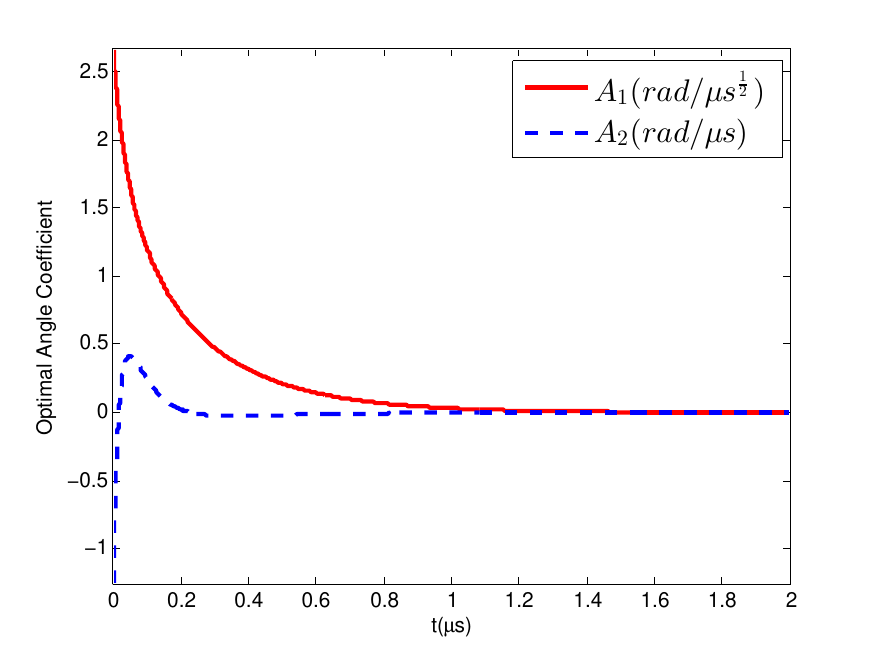}
\caption{
Optimal angle coefficients $A_1$ and $A_2$, evaluated as a function of time for the ASLO protocol for generation of the $N=3$ W state from the initial product state $|000\>$.
}
\label{fig:opt_angle_W}
\end{figure}

Experimentally it is generally easy to prepare qubits in a product state, so we shall assume without loss of generality that the initial state is $|000\>$. The advantage of keeping state evolution in the symmetric subspace is that this will help us rule out many unwanted states. This makes reaching the target state in the end more likely as well as finding the optimal angle numerically more efficient. We shall discuss this role of symmetry further when developing feedback protocols for the
GHZ state, where it places important physical constraints on the measurement observable. 
To numerically test the PaQS protocol, Eqs. \eqref{eqn:theta_parameterization} and \eqref{eqn:optimal_coefficient}, we start from the product state $|000\>$. Before evolving this state, we first locally rotate it with $U_F^W(\theta)$ to a state with maximum fidelity. Following this initial rotation, we evolve under the local optimization protocol, computing the optimal feedback coefficient at every time step using Eq.~\eqref{eqn:optimal_coefficient}. 

Fig.~\ref{fig:W_fidelity} shows the resulting trajectory ensemble average (TEA) fidelity (red solid line)  with respect to the W state, obtained by averaging over $1000$ trajectories. The TEA results are seen to saturate at fidelity $\sim 1$ at a time of $\sim 1.5\mu$s. 
We note that in order to ensure that the fidelity saturates at $1$, it is necessary to occasionally apply non-infinitesimal rotations when the second derivative test fails (Eq.~\eqref{eqn:2DTest}). 
Approximately 100 out of 10$^4$ trajectories, i.e., $\sim 1\%$ required at least one non-infinitesimal rotation angle (for a total of $\sim 10^5$ time steps). 
Without feedback, the fidelity remains constant at the value $0.44$ that is obtained after an optimal rotation of $\theta = 2\arcsin(\frac{1}{\sqrt{3}})$, which we plot for comparison with a 
blue dashed line.

We now compare the TEA-based protocol, which implicitly assumes knowledge of the entire measurement record for $\rho$ and requires real-time state estimation, with the more computationally and experimentally efficient ASLO protocol in which the feedback controls are optimized based on the average state $\bar{\rho}(t)$, Eq.~\eqref{eqn:ASLO_output_state}.
Fig.~\ref{fig:opt_angle_W} shows the evolution of the optimal feedback angle coefficients $A_1(t)$ and $A_2(t)$ evaluated by the ASLO protocol. 
The coefficient $A_2(t)$ determines the average value of $\theta^*$, 
while $A_1(t)$ determines its variance. It is evident that the average value is considerably smaller than 
the variance at all except the earliest times in the evolution.

Using the optimized ASLO feedback angles, Eq.~\eqref{eqn:optimal_angle}, 
with $\rho_t$ replaced by the unconditioned state $\bar{\rho_t}$ to control the averaged state dynamics gives rise to the fidelity shown in 
Fig. \ref{fig:W_fidelity} by the dashed red line. The same initial condition is used here as for the TEA protocol, \textit{i.e.}, the 
3-qubit product state $|000\>$. It is apparent that while the W state is reached with high fidelity within a comparable time of approximately 600 ns, the ASLO protocol nevertheless saturates slightly below unity, at $\sim 0.98$.
The origin of this difference lies in the different sampling of the density matrix that is enabled by the TEA and ASLO approaches.  Note that while the procedure for generating the coefficients $A_1(t)$ and $A_2(t)$ is identical, Eq.~\eqref{eqn:ASLO_output_state}, the input density matrices are different, with the TEA approach sampling these from many trajectories while the ASLO approach takes just one averaged trajectory. However, these two approaches can be identical in the situation where feedback cancels measurement noise exactly, as is the case for two-qubit optimal control \cite{Martin2017a}. 

It should be noted that although the ASLO protocol does not achieve unit fidelity, one can nevertheless still produce unit-fidelity states by adding a final projective measurement.  As the symmetry reduction has removed all degeneracy from the measurement operator $X$, the resulting measurement outcome now uniquely determines the state. Although the success probability under the ASLO protocol is less than one, it has been significantly enhanced by feedback. 
This argument applies to all protocols in which the symmetry reduction is made, so that one may thereby interpret the final ASLO fidelity as a success probability for perfect state preparation.

\subsection{Generation of general Dicke states}

We now consider the general Dicke states of Eq.~\eqref{eqn:Dicke} with $k >1$ and arbitrary $N$.  Several previous works have demonstrated deterministic Dicke state preparation with feedback, but these protocols were state-based and hence non-Markovian\cite{Thomsen2002,Stockton2004,Wei2015}. It is interesting to ask how well a Markovian protocol performs for the same task, particularly since the dynamical state estimation required for non-Markovian protocols becomes exponentially more challenging for larger systems. The most straightforward generalization of our $N=3$ qubit protocol is to simply add more operator components to $X_W$ and $H_F$
\begin{align}
    X_D^N &= \sigma_{z,1}+\sigma_{z,2}+\cdots + \sigma_{z,N} \\ \nonumber
    H_F^N &= \sigma_{y,1}+\sigma_{y,2}+\cdots + \sigma_{y,N}.
\end{align}
$H_F^N$ is still local as required. $X_D^N$ can be measured by placing each qubit in a separate cavity in an extension of the scheme demonstrated in \cite{Roch:2014ey}, or by coupling many qubits dispersively to the same cavity. The latter method (without continuous feedback) has been applied to generate spin squeezing in cold neutral atoms coupled to an optical cavity~\cite{Cox2016}. 

Imposing the permutation group symmetry on $N$ qubits allows the Dicke states to be represented within the symmetric subspace of dimension $N+1$ \cite{Stockton2003}. Each state in this subspace is just a Dicke state with $k$ excitations, which is also associated with a non-degenerate eigenspace of the observable $X_D^N$. In particular, the $W$ state for N qubits, defined as
\begin{equation}
|W\>=\frac{1}{\sqrt{N}}(|10\cdots 0\>_N+|01\cdots 0\>_N+\cdots +|00\cdots 1\>_N)
\end{equation}
belongs to the eigenspace of $X_D$ with eigenvalue $(N-1) - 1 = N-2$.
This implies that the computational resources required to compute the feedback protocol scale only polynomially with the number of qubits, which is a huge improvement compared to the exponential scaling of the full Hilbert space dimension. 
We note that other than for the smallest case of $N=2$ \cite{Martin2015,Martin2017a}, the average state $\bar{\rho}$ does not remain pure under ASLO feedback, so the stochastic terms do not cancel in Eq.~\eqref{eqn:WM} and the unaveraged dynamics (\textit{i.e.} conditioned on the entire measurement record) depends on the entire measurement record.

\begin{figure}[t]
\includegraphics[width=0.5\textwidth]{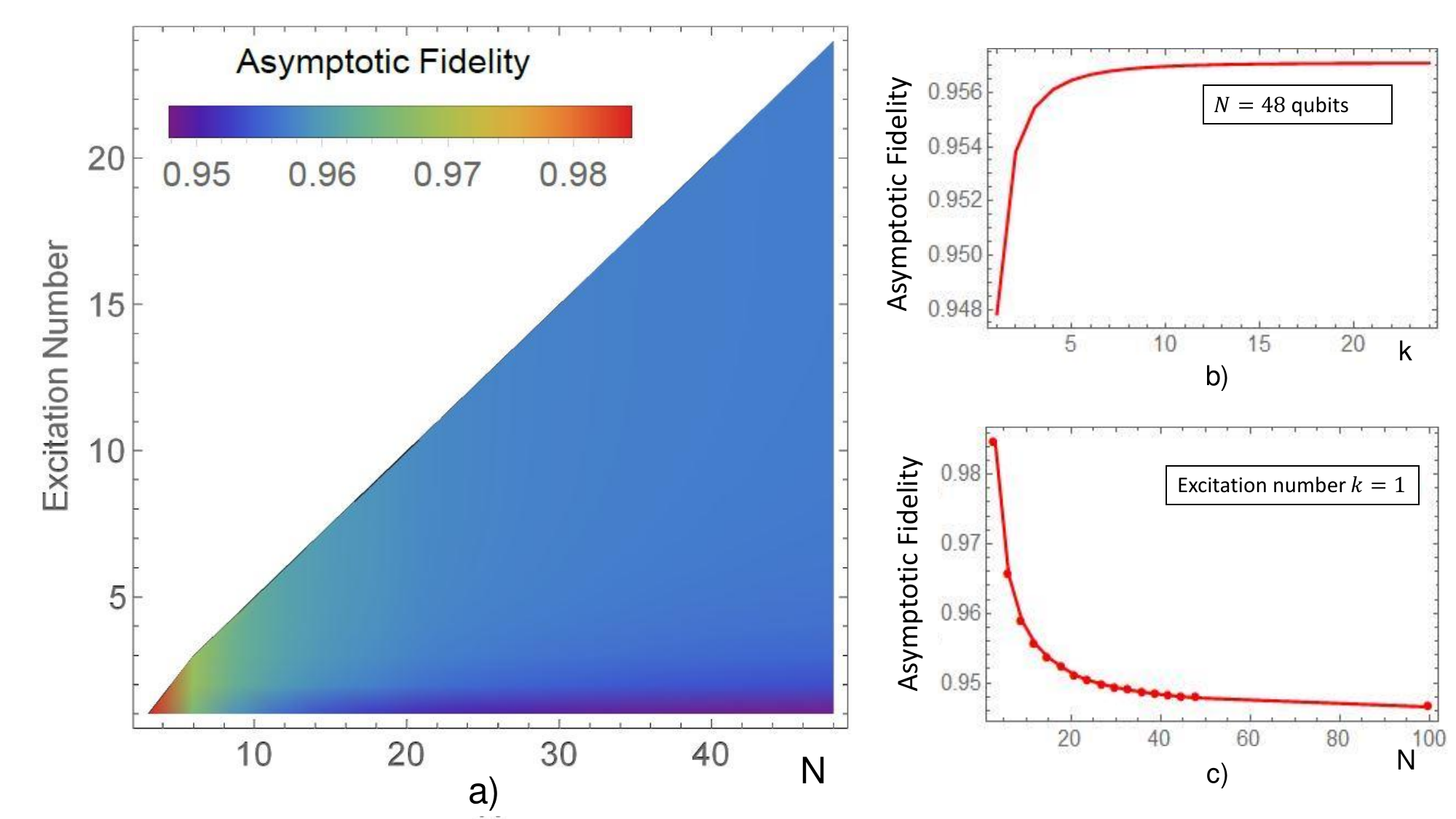}
\caption{Fidelity with respect to generalized Dicke states $|N,k\>$ 
under the ASLO protocol with locally optimal feedback, equations \eqref{eqn:ASLO_output_state} - \eqref{eqn:ASLOandUF}, with $U_F$ determined by the PaQS approach of section \ref{subsec:locallyoptimal}. Panel $a$) shows fidelity as a function of number of qubits $N$ (from $3$ to $48$, in increments of $3$) and excitation number $k$. 
The two right panels show cuts through this two-dimensional plot for fixed values of $N$ and $k$. 
Panel $b$): $k$-dependence of fidelity for $N=48$ qubits.  Panel $c$): $N$-dependence for states with $k=1$ excitation ($W$ states), extending to $N=100$ qubits.}
\label{fig:dicke_state}
\end{figure}

The results of ASLO calculations with the locally optimal protocol of section~\ref{subsec:locallyoptimal} (see Eq.~\eqref{eqn:WM}) for Dicke states with variable excitation number $k$ for up to $N=48$ qubits, are shown in Fig. \ref{fig:dicke_state}. 
We see that the final fidelities in Fig. \ref{fig:dicke_state} are all above 94\% despite the fact that the system does not take a predictable path through Hilbert space and hence an ASLO feedback controller does not know the true state. 
This is exemplified in the two inserts.  
The lower inset in Fig. \ref{fig:dicke_state} shows that the maximum fidelity for the W state ($k=1$) decreases with the number of qubits $N$.  This is not surprising, given the increased dimensionality of the system. However it is remarkable that the fidelity decreases only to $\simeq 0.94$ for $N=100$, indicating a strong robustness of the ASLO protocol for these target states.  The upper inset shows how the maximum fidelity depends on the excitation number $k$ for Dicke states of $N=48$ qubits. Here the fidelity increases with $k$, indicating that the more symmetric states, i.e., states that have closer to half of the qubits in $|0\rangle$ and half in $|1\rangle$, are more efficiently prepared.  Indeed we find empirically that $A_2=0$ for half-filled Dicke states with even $N$. 
The high performance thus appears to be due in part to the highly symmetric nature of the problem. The symmetry reduction to $|N,n\>$ has removed all degeneracy from the measurement operator $X_D^N$, so that the measurement outcome now uniquely determines the state.

\section{Generation of GHZ states}
\label{sec:GHZ}

The GHZ state
\begin{equation}\label{eqn:GHZ}
|\textrm{GHZ}\>=\frac{1}{\sqrt{2}}(|000\>+|111\>)
\end{equation}
is the three-qubit state with both maximal multi-particle entanglement and maximal fragility of three-way entanglement~\cite{Dur2002,Gisin1998}.  
The state 
\begin{equation}
|\textrm{GHZ}\>_N=\frac{1}{\sqrt{2}}(|00\cdots 0\>_N+|11\cdots 1\>_N).
\end{equation} 
is the extension to $N$ qubits.  We shall use the notation $|\textrm{GHZ}\>$ to refer specifically to the $N=3$ state.

Generation of GHZ states using the above method poses new challenges. As noted above, measurement operators that are linear combinations of single-body observables like $Z$ are the easiest to implement in remote and weakly interacting settings. The GHZ state is an eigenvector of such operators, with the general form
\begin{align} \label{eqn:GHZ_observable_less_sym}
   X_G = (a+b)Z_1 - a Z_2 - b Z_3.
\end{align}
However, such an operator does not satisfy the full permutation symmetry of the target state. Furthermore, the target state $|\textrm{GHZ}\>$ is now one of a degenerate set of eigenstates of $X_G$. As we saw for the W state above, it is important that  the target state constitute a non-degenerate eigenspace within a subspace that is determined by a symmetry imposed by the measurement operator. In order to achieve this, we first need to impose the correct symmetry requirement on the measurement operator.

We consider here two forms of measurement operators for the GHZ state.  Both of these possess the full $S_3$ permutation symmetry 
and the additional bit-flip symmetry that characterizes 
the GHZ state. The first approach uses a symmetrized sum of two-qubit operators that imposes the additional bit flip symmetry of the GHZ state. The second approach
achieves the same symmetries by the action of 
multiple measurement operators with single-qubit components.

\subsection{Two-qubit measurement operators} 
\label{subsec:2qubitGHZ}
To prepare the GHZ state we can use the following symmetric two-qubit measurement observable
\begin{equation}\label{eqn:observable_GHZ}
 X_{G}^{S}=\sigma_{z,1}\sigma_{z,2}+ \sigma_{z,2} \sigma_{z,3} +\sigma_{z,3}\sigma_{z,1}.
\end{equation}
$X_{G}^{S}$ is the lowest order $n$-qubit observable that possesses both full permutation symmetry and bit-flip symmetry. Together, these two symmetries induce a symmetric subspace $\mathcal{H}_{G}^S$ of the full Hilbert space.
The basis vectors of $\mathcal{H}_{G}^S$ are expressed in the computational basis as 
\begin{align}
|\phi_{G}^1\> &= \frac{1}{\sqrt{2}}(|000\> + |111\>)\\
|\phi_{G}^2\> &= 
\frac{1}{\sqrt{6}}(|011\> + |101\> + |110\> + |100\> + |010\> + |001\>).
\end{align}
$|\phi_{G}^i\>, i=1,2$ are eigenstates of $X_{G}^{S}$ with eigenvalues $e_1=+3$ and $e_2=-1$.  Within this subspace, the GHZ state $|\phi_{G}^1\>$ is then a non-degenerate eigenstate of the measurement operator $X_{G}^{S}$. Thus, restricting the dynamics governed by Eq.~\eqref{eqn: output_state} to this subspace and employing a cost function that is biased towards the state $|\phi_{G}^1\>$, will allow continuous measurement of $X_{G}^{S}$ together with the feedback controls to extract the GHZ state. 
Similarly to the procedure for the Dicke states in the previous section, we achieve this by using the fidelity with the target state as the cost function, i.e., $\mathcal{F}_G =  \<\phi_G^1|\rho_t|\psi_G^1\rangle$.  But in contrast to the imposition of only permutation symmetry on the feedback operator in section \ref{sec:W}, we need to now impose the full permutation and bit-flip symmetry of $X_{G}^{S}$ on the feedback rotation operator $U_F$, to ensure that the action of this does not take the state out of $\mathcal{H}_{G}^S$.  Therefore we define $U_F$ here by a single rotation around the $x$ axis, which is consistent with the bit-flip symmetry, and set the rotation angles to be equal for all three qubits, to be consistent with the permutation symmetry. This yields the rotation operator
\begin{equation}\label{eqn:control_op_GHZ}
 U_F^{G}(\theta) = e^{-i\frac{\theta}{2}(\sigma_{x,1}+\sigma_{x,2}+\sigma_{x,3})}.
\end{equation}
If the initial state is within $\mathcal{H}_G^S$, then the combined action of $U_F^{G}(\theta)$ and $X_G^S$ will ensure that the subsequent evolution always remains in $\mathcal{H}_G^S$. 

Similar to the analysis above for the $W$ state, we then determine the locally optimal angle $\theta^*$ by maximizing the fidelity at each time step 
\begin{equation}
\mathcal{F}_G(t+dt) = \<\textrm{GHZ}|\rho_{t+dt}^c|\textrm{GHZ}\>,
\end{equation}
where $\rho_{t+dt}^c$ is again the output state given by Eq.~\eqref{eqn: output_state}. 
When the symmetric feedback operator \eqref{eqn:control_op_GHZ} is used with the GHZ state target, we can obtain an exact solution for $\partial \mathcal{F}/\partial \theta$  and $\partial^2 \mathcal{F}/\partial \theta^2$ and hence find the explicit values for the optimal angle $\theta^*$.
Full details of the solution are given in appendix~\ref{app:GHZ}, where we show that the optimal angle is always either $0$ or $\pi/2$, with the specific choice given by the sign of a state-dependent term. Thus there is no need to use a PaQS protocol based on infinitesimal rotation angles in this situation. This also means that we cannot use the SME Eq.~\eqref{eqn:WM}, which assumes infinitesimal rotations, to simulate the dynamics and instead we must use the full POVM equation, Eq.~\eqref{eqn: output_state}. 

The GHZ state also allows an interesting alternative optimization approach, deriving from the fact that after the dynamics are constrained to the symmetric subspace, the dimension of the Hilbert space is reduced from eight to two. Consequently, under these constrained dynamics the three-qubit problem can be mapped to a single qubit problem. This mapping is described explicitly in appendix~\ref{sec:single_qubit}, where it is shown that this allows an alternative approach to determination of the optimal angle that also results in an optimal angle of either $0$ or $\frac{\pi}{2}$. 

\begin{figure}[t]
\includegraphics[width=0.5\textwidth]{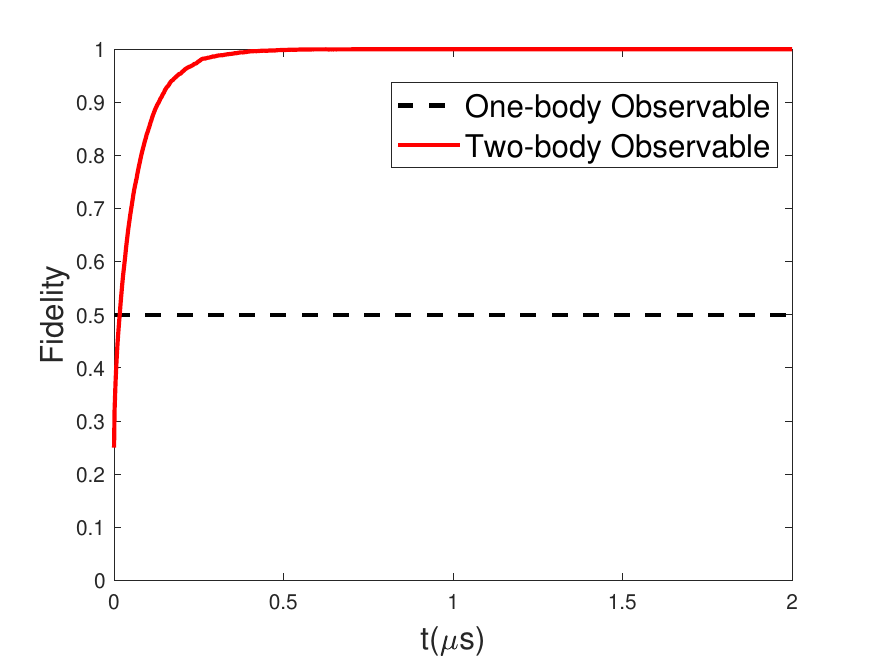}
\caption{Fidelity $\mathcal{F}_G(t)$  of the controlled state with respect to GHZ state as a function of time, obtained from  the locally optimal ASLO approach. The red line is the result obtained with the symmetric two-body measurement operator $X_{G}^{S}$. The blue line shows the result obtained with the non-symmetric one-body measurement operator $X_{G}$.  All simulations were run with measurement strength $k=1MHz$ and time step $\Delta t \ll \frac{1}{k}$, starting from the complete superposition state (see text).}
\label{fig:GHZ_fidelity}
\end{figure}

Fig. \ref{fig:GHZ_fidelity} shows the time dependence of the fidelity of formation of the GHZ state obtained using the locally optimal protocol within the ASLO approach, with the symmetrized measurement $X_{G}^{S}$, and using as initial condition the complete superposition state  
\begin{equation}\label{eq:fullsuperposition}
\frac{1}{\left(\sqrt{2}\right)^3}\left(|0\> + |1\> \right)\otimes \left(|0\> + |1\> \right)\otimes \left(|0\> + |1\> \right).
\end{equation}
We see that in this situation the fidelity under the ASLO protocol asymptotically approaches unity. 

It is instructive to compare the performance of this protocol with a symmetrized two-body measurement operator, to that obtained from feedback control based on measurements of an observable not respecting this permutation symmetry. We consider here the one-body observable of Eq.~\eqref{eqn:GHZ_observable_less_sym}
with $a = b = 1$, which is permutation symmetric only with respect to exchange of the last two qubits. The target state is then contained in a degenerate eigenspace with eigenvalue zero that is spanned by $\frac{1}{\sqrt{2}} (|000\> \pm |111\>)$. 
While we can still impose the symmetry condition on the feedback rotation operator, the measurement is now unable to distinguish the target GHZ state from another state ($\frac{1}{\sqrt{2}}(|000\> - |111\>)$) within the degenerate eigenspace. We therefore expect that any feedback protocol based on this measurement will be unlikely to reach unit fidelity. In fact, the results achieved with this protocol, shown as the blue line in Fig. \ref{fig:GHZ_fidelity}, are of very low quality.

\begin{figure}[ht]
\includegraphics[width=0.5\textwidth]{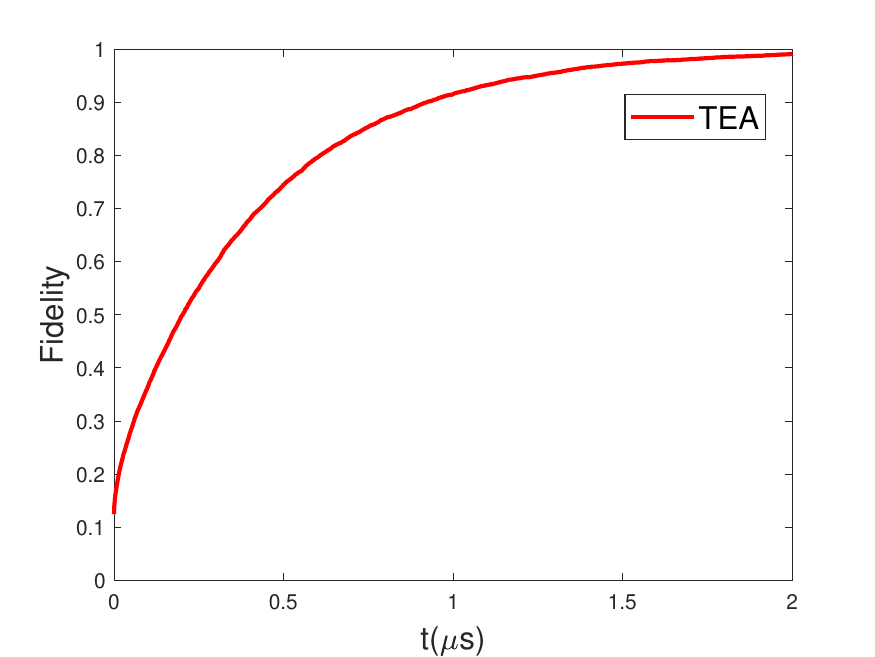}
\caption{Fidelity of four-qubit GHZ state, computed with the TEA protocol using the complete superposition as initial state.}
\label{fig:N=4GHZ}
\end{figure}

We can also generalize our protocol to the GHZ state of $N$ qubits, which is defined as
\begin{equation}
|\textrm{GHZ}\>_N=\frac{1}{\sqrt{2}}(|00\cdots 0\>_N+|11\cdots 1\>_N).
\end{equation}
The observable we use is still the two-body symmetrized observable
\begin{equation}
 X^S_{G}=\underset{i<j}{\Sigma}\sigma_{z,i}\sigma_{z,j}.
\end{equation}
After imposing both the permutation and bit flip symmetry, the corresponding symmetric subspace $\mathcal{H}_{G}^S$ now has dimension $\frac{N+1}{2}$ when $N$ is odd, and dimension $\frac{N}{2}+1$ when $N$ is even. Explicitly, the symmetrized basis set is given by
\begin{align}
|\phi_{G}^1\> &= \frac{1}{\sqrt{2}}(|00\cdots0\>_N + |11\cdots1\>_N)\\
|\phi_{G}^2\> &= \frac{1}{\sqrt{2N}}\Sigma_{P\in S_N}P(1 + \Pi)|10\cdots0\>_N\\
              &\vdotswithin{=} \notag \\
|\phi_{G}^m\> &= \frac{1}{\sqrt{2\binom{N}{m}}}\Sigma_{P\in S_N}P(1+\Pi) |\underbrace{11\cdots1}_m 0\cdots0\>_N   
\end{align}
where $\Pi = X_1X_2\cdots X_N$ is the $N$-qubit bit flip operator. The index $m$ indicates how many spins are pointing downward and runs from $0$ to either $\frac{N-1}{2}$ ($N$ odd) or $\frac{N}{2}$ ($N$ even). Note that every basis vector $|\phi_G^m\>$ is a non-degenerate eigenstate of the observable $X_G^S$. In particular, the GHZ state is just $|\phi_G^1\>$, with eigenvalue $\binom{N}{2}$. Following the same procedure as for $N=3$, we choose the feedback rotation operator to be of the form $U_F(\theta)=\bigotimes\limits_{i=1}^N e^{-i\frac{\theta}{2}X_i}$ and start from the symmetric initial state $|\psi(0)\>=\bigotimes\limits_{i=1}^N\frac{|0\>+|1\>}{\sqrt{2}}$. 
The locally optimal rotation angles are calculated using the same local expansion approach employed for $N=3$ qubits (see appendix~\ref{app:GHZ_1}).

Results obtained with the trajectory ensemble approach for generation of the $N=4$ GHZ state using this locally optimal GHZ protocol are shown in Fig.~\ref{fig:N=4GHZ}.
Undertaking these non-Markovian locally optimal calculations for the GHZ state is significantly more expensive than the corresponding calculations for the W state in section~\ref{sec:W}. While possible in principle, this again motivates the use of the ASLO approach for larger $N$ values, as was done for the generalized Dicke states for large $N$ in section~\ref{sec:W}. In the following subsection we address this, together with the goal of finding a feedback protocol for GHZ states that also uses only single qubit measurement and feedback Hamiltonian operators. To do this, we have to first generalize our PaQS protocol to allow multiple operators acting on each qubit.

\subsection{Multiple measurement and multiple feedback Hamiltonian operators}
\label{subsec:multioperatorGHZ}

The extension of PaQS to handle multiple simultaneous measurement operators and feedback Hamiltonians is straightforward so long as the feedback Hamiltonians commute. In what follows, we extend PaQS to cover these cases. The result also turns out to hold for non-commuting Hamiltonians under certain conditions~\cite{martin2019single}.  We then show that with this generalization it is possible satisfy the permutation symmetry of the GHZ state, while also restricting the measurements to the experimentally more accessible linear combinations of  single-body observations. We start by noting that the GHZ state is also an eigenstate of the following operators
\begin{align}
\label{eq:M_ij}
 M_{ij} = \frac{\sigma_{z,i} - \sigma_{z,j}}{2},
   ~ i,j=1\cdots N,~i\neq j.
\end{align}
These resemble the half-parity measurement operators of Table~\ref{tab:PaQSTable}, with the crucial difference that  $M_{ij}$ now give identical outcomes on the states $|00\>$, $|11\>$ rather than on the states $|01\>$, $|10\>$ of each $i,j$ pair of qubits. The action of this set of measurement operators imposes both permutation and bit flip symmetry on the corresponding set of terms in the SME, Eq.~\eqref{eqn:measurement}.

To find the locally optimal protocol, we need to generalize the expression for $\theta_\text{opt.}$ to handle multiple simultaneous measurements $M_i$ and multiple feedback operations $H_j$. In general there is no preferred pairing between the measurement and feedback operations in these groups, so that the $i$th measurement outcome may affect how we apply the $j$th feedback Hamiltonian. This forces us to rederive the feedback master equation with a more general feedback unitary,
\begin{widetext}
\begin{align} \label{eqn:generalfeedback}
	U &= \exp\left( -i \sum_j H_j \theta_j \right) \\ \nonumber
    &= I - i\sum_{ij} A_{ij} H_j dW_i - \left[ i\sum_j B_j H_j + \frac{1}{2} \sum_{ijk} A_{ij} A_{ik} H_j H_k \right] dt \\ \nonumber
    \theta_j &= B_j dt + \sum_{i} A_{ij} dW_i,
\end{align}
\end{widetext}
where $A_{ij}$ and $B_i$ are analogous to $A_1$ and $A_2$ respectively, and our goal is now to find the locally optimal values of all $\{A_{ij}\}$ and $\{B_i\}$. The presence of cross terms in Eq.~\eqref{eqn:generalfeedback}  modifies the resulting feedback master equation
\begin{widetext}
\begin{align} \label{eqn:MultiPaQSSME}
  \rho(t+dt) &= \rho -i \sum_j B_j [H_j,\rho] dt + \sum_i \Big[{2k} \mathcal{D}[M_i]\rho ~ dt + \sqrt{\eta_i{2k}}\mathcal{H}[M_i]\rho~ dW_i - i \sum_{j} A_{ij} [H_j, \rho] dW_i  \nonumber \\ 
  -i \sum_j & \sqrt{\eta_i} A_{ij}[H_j, {\sqrt{2k}}(M_i \rho + \rho M_i^{\dagger})]dt + \sum_{jk} A_{ij} A_{ik} [H_k \rho H_j - \frac{1}{2}(H_j H_k \rho + \rho H_j H_k) ]dt \Big]
\end{align}
\end{widetext}
In contrast to Eq.~\eqref{eqn:WM}, the last term is not in Lindblad form due to the presence of cross terms, though it may be cast into Lindblad form by defining the modified feedback operators $\tilde{H}_i = \sum_j A_{ij} H_j$. For simplicity, we assume that the control Hamiltonians commute pairwise, although the end result is essentially unmodified if one relaxes this assumption~\cite{martin2019single}.\footnote{In general, one must have that the $H_i$ form a Lie algebra \textit{i.e.}, that the vector space formed by $H_i$ is closed under commutation.} The locally optimal feedback coefficients must satisfy $\<\psi_T|[H_\alpha, d\rho]|\psi_T\>=0$, this time for all $\alpha$. Dealing first with the $dW$ terms, we find
\begin{align}
\sum_i & \Bigg[ \sum_j -i A_{ij} \underbrace{\<\psi_T|[H_\alpha,[H_j,\rho]]|\psi_T\>}_{c_{j\alpha}} \\ \nonumber
&+ \underbrace{\sqrt{\eta_i}\<\psi_T|[H_\alpha, {\sqrt{2k}}(M_i\rho + \rho M_i^\t)]|\psi_T\>}_{a_{i\alpha}} \Bigg] dW_i = 0
\end{align}
The solution is evident if we rewrite the expression in matrix form with the help of $a$ and $c$, which are state-dependent
\begin{align} \label{eq:MultiPaQSA}
-i A c + a = 0 ~~ \implies A = -i a c^{-1}.
\end{align}
In general, one should use the Moore-Penrose pseudoinverse to handle the case in which $c$ is not invertible\cite{martin2019single}. With a solution for $A$ in hand, we can solve for the $B$ coefficients by collecting $\mathcal{O}(dt)$ terms from $\<\psi_T|[H_\alpha, d\rho]|\psi_T\>=0$ 
\begin{align} \label{eq:MultiPaQSB}
&b_\alpha \equiv \<\psi_T|[H_\alpha,\sum_i \Bigg[ \mathcal{D}[M_i]\rho -i \sum_j \sqrt{\eta_i}A_{ij}[H_j, {\sqrt{2k}}(M_i\rho+\rho M_i^\t)] \\ \nonumber
&~~~~~~+ \sum_{jk}A_{ij}A_{ik} [H_k\rho H_j - \frac{1}{2}(H_j H_k \rho + \rho H_j H_k)]\Bigg]]|\psi_T\> \\ \nonumber
&-i\sum_i B_i c_{i\alpha} + b_\alpha = 0 ~~\implies \vec{B} = -i\vec{b}c^{-1}
\end{align}
treating $\vec{b}$ and $\vec{B}$ as row vectors. Eqs.~ \eqref{eq:MultiPaQSA} and~\eqref{eq:MultiPaQSB} generalize Eq.~\eqref{eqn:optimal_coefficient}.

Now we have a locally optimal solution for general measurements and feedback. To generate the GHZ states, we will set the measurement operators $M_i$ to be $M_{ij}$ as given in Eq.~\eqref{eq:M_ij}, which ensures that the measurement term in the SME Eq.~\eqref{eqn:MultiPaQSSME} satisfies both permutation and bit-flip symmetries. Then we still need to choose a set of feedback operators. For simplicity, we limit ourselves to local $\sigma_y$ rotations. $c$ can fail to be invertible if there are to many $H_i$s, and we have learned the importance of symmetry, so some care is required in choosing them. One obvious choice of Hamiltonian basis is $H_i = \sigma_{y,i}$, one for each qubit. However, this basis does not enforce the $|11...1\> \leftrightarrow |00...0\>$ bit flip symmetry of the GHZ state and the $M_{ij}$ operators. A better choice would be a set orthogonal to $\sum_i \sigma_{y,i}$, such as
\begin{align} \label{eqn:GHZHFs1}
    H_i = \frac{\sigma_{y,i}-\sigma_{y,i+1}}{2}, ~~~
    i=1...N-1
\end{align}
where $N$ is the number of qubits. Although this basis appears to break permutation symmetry, any Hamiltonian of the form 
\begin{align} \label{eqn:GHZHFs2}
H_{ij} = \frac{\sigma_{y,i}-\sigma_{y,j}}{2}
\end{align}
can be written as a linear combination of the set of $H_i$. Thus the locally optimal solution can retain permutation symmetry.

\begin{figure*}[ht]
\centering
\includegraphics[width = 1\textwidth]{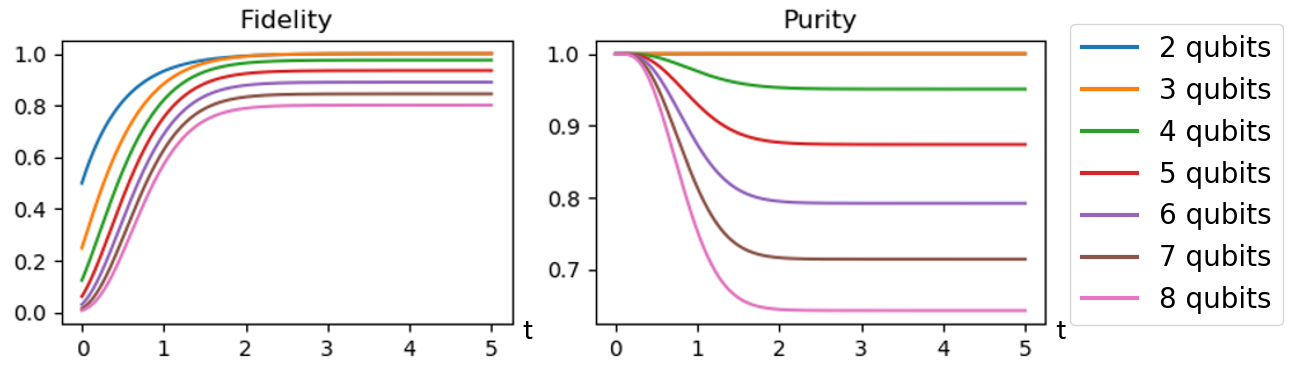}
\caption{Fidelity and purity of the controlled state under the multiple measurement and multiple feedback operator protocol for generation of $N$ qubit GHZ states, in the ASLO implementation. For $N=2$ and $N=3$ the purity remains exactly $1$ for all time.}
\label{fig:GHZFidelity}
\end{figure*}

The time dependence of GHZ state generation under ASLO feedback is plotted in Fig.~\ref{fig:GHZFidelity}. The 2-qubit case reproduces the half-parity measurement of Refs.~\cite{Roch:2014ey} and~\cite{Martin2015} under the transformation $\sigma_{z,2} \leftrightarrow -\sigma_{z,2}$, while the remaining results $N=3$ through $N=8$ are novel. Although we use Eq.~\eqref{eqn:GHZHFs1} as our feedback Hamiltonians, the locally optimal protocol takes linear combinations of the form Eq.~\eqref{eqn:GHZHFs2}, so that each $H_{ij}$ is paired with its corresponding $M_{ij}$. We speed up the numerics by taking this observation into account, so that many terms in Eq.~\eqref{eqn:MultiPaQSSME} may be set to zero. Perhaps most interestingly, the $N=3$ case reaches exactly unit fidelity, and the purity of the average state remains 1 throughout the feedback process, as in the example at the beginning of this section. One can derive an analytic solution in this case.

\subsection{Feedback protocols for GHZ states based on non-fidelity measures}
As noted in section~\ref{subsec:locallyoptimal} above, it is also possible to construct feedback protocols based on cost functions providing a direct measure of entanglement, rather using as cost function the fidelity with a specific target state.  While, as seen above, the latter choice simplifies many calculations, it does however require that a single specific target state is singled out. This ignores the possibility that one might be able to do better by targeting a different target state that is nevertheless locally equivalent to the desired target state.
In appendix~\ref{app:non-markovian} we present a method for deriving locally optimal protocols based on entanglement measures, and apply it to directly optimize the three-tangle, a measure of entanglement for three qubits that achieves its maximal unit value for a GHZ state and all locally equivalent states.

\section{Conditions for deterministic evolution under measurement-based feedback control} \label{sec:deterministic}
This work has revealed a number of intriguing examples in which feedback perfectly eliminates the randomness of the measurement process. These cases are highlighted with asterisks in table \ref{tab:PaQSTable}. All of these specific protocols can be shown to be globally optimal for the task at hand, except for the 3-qubit GHZ protocol that we conjecture to also be optimal. These examples also yield analytic solutions in all cases, further indicating that they belong to a natural class of protocols in continuous measurement.  Ref.~\cite{Martin2017a} provides detailed analysis of this for the half- and full-parity Bell state measurements in table \ref{tab:PaQSTable}.

It is straightforward to introduce a strict necessary condition for whether or not deterministic evolution is possible, given fixed measurement and feedback operators. This condition also provides a simple tool for finding solutions to the general proportional 
feedback master equation, supposing
that we apply proportional feedback of the form $U=\exp(-i H(P dr+c dt))$, where $H$ is the feedback operator, $P$ and $c$ are time-dependent scalars, and $dr = \sqrt{\eta}\<M+M^\dagger\> + dW$ is the incremental output signal for the measurement operator $M$.\footnote{Note that we have absorbed the measurement strength into $M$ here.} 
Applying $U$ to the stochastic Schr\"odinger equation yields a pure-state equation of motion for $|\psi\>$ valid for $\eta=1$
\begin{widetext}
\begin{align}
    |\psi(t+dt)\> = \Big[I&+(M-m-iPH)dW - \frac{1}{2}(M^\t M-2M m+m^2+P^2 H^2)dt \\ \nonumber
    &-i H(P(M+m)+c)dt \Big] |\psi(t)\>
\end{align}
\end{widetext}
where $m\equiv \<M+M^\t\>/2$. We should allow for the solution to be deterministic up to a global phase, so we make the substitution
\begin{align}
    H \rightarrow H+V_0.
\end{align}
For a deterministic solution, the $dW$ terms of the stochastic Schr\"odinger equation must cancel, \textit{i.e.},
\begin{align} \label{eqn:DetdWCondition}
    [M-m-iPH-iPV_0]|\psi\> = 0.
\end{align}
This is a nonlinear constraint on $|\psi\>$, since the expectation value $m$ depends on the state. It can only be satisfied if $|\psi\>$ is a right eigenvector of $M-i P H$. To see that this necessary condition is also sufficient, compute the corresponding eigenvalue $\lambda$ by taking an expectation value
\begin{align}
    \<\psi|&[M-i P H] |\psi\> = \lambda = \<M\>-i P\<H\>
\end{align}
and then compute the left-hand side of \eqref{eqn:DetdWCondition}
\begin{widetext}
\begin{align}
[M-m-iPH-iPV_0]|\psi\> &= \Big[\<M\>-iP\<H\>-\frac{\<M+M^\t\>}{2}-iP V_0\Big]|\psi\> \\ \nonumber
&=\Big[\frac{\<M^\t-M\>}{2}-i P \<H\> - i P V_0\Big]|\psi\>
\end{align}
\end{widetext}
As $\<M^\t-M\>$ is purely imaginary, we can always find a $V_0$ such that the above expression is zero. Thus feedback can exactly cancel the measurement-induced noise on $|\psi\>$ if and only if $|\psi(t)\>$ is a right eigenvector of $M-i P H$. As $P$ is a free parameter, we can check if this condition is possible for any value of it.

With the stochastic terms cancelled, the next step is to see if the deterministic part of the equations of motion can maintain the above condition for a finite time interval. Given the above result, we can search for such solutions within each eigenspace of $M-i P H$ separately (the eigenspaces will typically evolve continuously as a function of $P$).  Let $\{|v_i(P(t))\>\}$ be a set of orthonormal eigenvectors associated with a chosen eigenspace. To obtain a deterministic solution, we must have
\begin{align}
    |\psi(t)\> = \sum_i c_i(t) |v_i\>.
\end{align}
Now consider the deterministic part of the equation of motion. Using the above expansion, we have
\begin{widetext}
\begin{align}
    \frac{d|\psi\>}{dt} &= \sum_i \dot{c}_i(t)|v_i\>+c_i(t) \frac{dP}{dt}\frac{d|v_i\>}{dP} \\ \nonumber
    &= \left[\frac{1}{2}(M^\t M-2M m+m^2+P^2 H^2)dt - i H(P(M+m)+c)dt\right]|\psi(t)\>.
\end{align}
\end{widetext}
As $m=\text{Re}(\lambda)$ and $\lambda(t)$ is fixed by working within a fixed eigenspace, the above equation of motion is linear in $|\psi\>$, which significantly simplifies the finding of a solution. One can use the above procedure to identify candidate solutions, which appear to be plentiful and relatively unexplored. For example, the multiple measurement, multiple feedback operator GHZ state protocol in section~\ref{eqn:measurementoutcome} yields an analytic solution for $N=2$ and $N=3$ under the above procedure.

\section{Discussion and Conclusions}\label{sec:con}
We have presented a general, analytic construction of locally optimal measurement-based feedback protocols, termed PaQS, in which the feedback unitary has two independently varying components, one having dependence on the measurement outcome and the other having dependence only on the quantum state.
We showed that the resulting protocol can be modified to generate an average state locally optimal (ASLO), or Markovian protocol that can be efficiently implemented. 

In this work we demonstrated the effectiveness of PaQS feedback by applying it to the generation of W, general Dicke and GHZ states for $N = 3 - 100$ qubits, highlighting the utility of symmetry constraints. The ASLO versions of these protocols reach target state fidelities above 94\%, despite the significant constraints imposed by Markovianity.  This shows that a simple Markovian-type feedback strategy exists to prepare any of these maximally spin-squeezed states deterministically.

While we have focused on application to entanglement generalization, the full scope of Eqs. (\ref{eqn:MultiPaQSSME})-(\ref{eq:MultiPaQSB}) extends to virtually any application of measurement-based feedback. The ability of these equations to reproduce all existing feedback protocols that we have tried them on (see table \ref{tab:PaQSTable}) and the ease with which they generate new protocols suggests that there is much more to be gained from them. 
One immediate extension of this work would be to study the behavior under non-ideal measurement conditions, i.e., for $\eta < 1$.

Several interesting challenges remain regarding systematization of this PaQS feedback approach. Although we have so far only considered specific target states, one may ask more generally which classes of quantum states may be generated under constraints on the measurement and feedback operators. Similar questions have been successfully answered in the context of Markovian stabilization \cite{johnson2017exact,ticozzi2018alternating} and it would be natural to apply and extend these questions to non-Markovian situations. 
Another important direction for future applied work would be to classify what states can be prepared deterministically (under a suitable definition) using local feedback and measurement operators that are linear combinations of local observables.
Alternatively, one could ask the converse question of how to efficiently determine suitable measurement and feedback operators given a desired target state. Finally, questions of state stabilization and robustness leave important directions for future work. Recent work has shown that global exponential stabilization of two-level systems is possible  using a quantum state and proportional control method similar to that introduced here~\cite{Cardona2018}. Whether PaQS yields global or exponential stabilization  for higher dimensional quantum systems is an important question for further work.

\begin{acknowledgements}
This work was supported by
Laboratory Directed Research and Development (LDRD)
funding from Lawrence Berkeley National Laboratory,
provided by the U.S. Department of Energy, Office of
Science under Contract No. DE-AC02-05CH11231. 
It was also partially supported by the Office of
the Director of National Intelligence (ODNI), Intelligence Advanced
Research Projects Activity (IARPA), via the U.S. Army Research Office
contract W911NF-17-C-0050. The views and conclusions contained herein are
those of the authors and should not be interpreted as necessarily
representing the official policies or endorsements, either expressed or
implied, of the ODNI, IARPA, or the U.S. Government. The U.S. Government
is authorized to reproduce and distribute reprints for Governmental
purposes notwithstanding any copyright annotation thereon. The effort of LM was supported by grants from the National Science Foundation Grant No. (1106400) and the Berkeley Fellowship for Graduate Study.

$^\dagger$ S. Z. and L. S. M. contributed equally to this work.
\end{acknowledgements}

\begin{appendix}
\section{\label{app:2DTest} Second Derivative Test for Local Optimal Control}

In this appendix we discuss the situation when the infinitesimal solution Eq.~\eqref{eqn:theta_parameterization} with Eq.~\eqref{eqn:optimal_coefficient} fails to pass the second derivative condition, Eq.~\eqref{eqn:2DTest}. Using the same expansion as in section~\ref{subsec:locallyoptimal}, we have
\begin{widetext}
\begin{align}
\<\psi_T|[H,[H,\rho^c]]|\psi_T\> & = \<\psi_T|[H,[H,\rho]]|\psi_T\> + \<\psi_T|[H,[H,{\sqrt{\eta}} (Y\rho+\rho Y^\dagger)]] -i A_1[H,[H,[H,\rho]]]|\psi_T\>dW \\
 & + \<\psi_T|[H,[H,\mathcal{D}[Y]\rho]]+A_1^2[H,[H,\mathcal{D}[H]\rho]]-i [H,[H,[H, {\sqrt{\eta}} A_1 (Y\rho +\rho Y^\dagger) + A_2\rho]]]|\psi_T\>dt > 0
\end{align}
\end{widetext}
We can see that due to the stochastic nature of $dW$, there will always be some non-zero probability that this condition is violated. Specifically, when
\begin{widetext}
\begin{equation}
|dW| > \bigg|\frac{\<\psi_T|[H,[H,\rho]]|\psi_T\> + \<\psi_T|[H,[H,\mathcal{D}[Y]\rho]]+A_1^2[H,[H,\mathcal{D}[H]\rho]]-i [H,[H,[H, {\sqrt{\eta}} A_1 (Y\rho +\rho Y^\dagger) + A_2\rho]]]|\psi_T\>dt}{\<\psi_T|[H,[H,{ \sqrt{\eta}}(Y\rho+\rho Y^\dagger)]] -i A_1[H,[H,[H,\rho]]]|\psi_T\>}\bigg|
\end{equation}
\end{widetext}

the infinitesimal solution will give a local minimum instead of a maximum. As long as the term $\<\psi_T|[H,[H,\rho]]|\psi_T\>$ is not zero (note that this term is always non-negative if the fidelity of $\rho$ is assumed to be locally maximal), in the limit that $dt$ goes to zero, the violation probability will be small for sufficiently small dt.

\section{Optimal Angle for GHZ State with Two-body Observable}
\label{app:GHZ}
\subsection{Local Expansion Method}
\label{app:GHZ_1}
We can also use the calculus method of local expansion employed in section \ref{sec:cont_meas} in order to find the optimal angle for the GHZ state. Using Eq.~\eqref{eqn:optimal_condition}, we have an explicit expression of $\mathcal{G}$ proportional to $\sin(2\theta)$ provided the input state is an extremal state. So the optimal angle would be either $0$ or $\frac{\pi}{2}$.
However, an extremal state can be either a local maximum or minimum. To determine which of these will be generated at time $t+dt$, we look again at the second derivative of the fidelity function (Eq.~\eqref{eqn:2DTest}):
\begin{align}
\frac{d^2 \mathcal{F}}{d\theta^2}&=\cos(2\theta)[\frac{3-6\rho_{11}-2\sqrt{3}\rho_{12}}{2}\\\nonumber
&+4\sqrt{2k}dW(6\rho_{11}^2-\sqrt{3}\rho_{12}+2\rho_{11}(-3+\sqrt{3}\rho_{12}))\\\nonumber
&+16\sqrt{3}k\rho_{12}dt].
\end{align}
It is then evident that the optimal choices of angle to ensure that the sign of the second derivative is negative are $\theta^* = 0$ or $\theta^* = \pi/2$, depending on the sign of the state-dependent term in parentheses.

\subsection{Mapping of $N=3$ GHZ state to a single qubit state}\label{sec:single_qubit}
\begin{figure}[t]
\includegraphics[trim = 10cm 3cm 10cm 3cm, clip=true, width=0.5\textwidth]{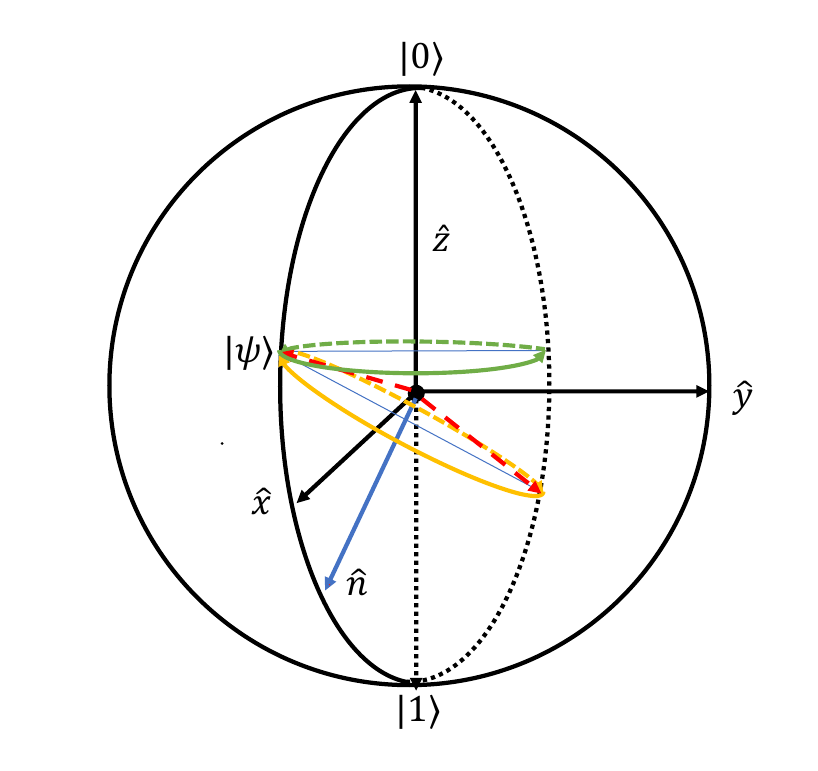}
\caption{
Bloch sphere illustrating the geometric meaning of the locally optimal control protocol for the GHZ state. The blue vector $\hat{n}=(\frac{\sqrt 3}{2}, 0, \frac{1}{2})$ is the rotation axis. Initially, our state is located exactly along this direction. Note that the measurement will always keep the state in the $xoz$ plane, where $o$ denotes the origin. Suppose that after a measurement, the state is one of the two red dashed vectors. The feedback operator will rotate this vector along the yellow circle around the $\hat{n}$ axis. To maximize the fidelity with respect to the $|\tilde{0}\>$ state, we should always rotate it to the {\it upper} position labeled by $|\psi\>$. This means that the rotation angle $\theta$ should be $0$ or $\frac{\pi}{2}$, depending on whether the state is above or below $\hat{n}$ after the measurement. The green circle at constant $z$ is drawn for reference as a visual aid.}
\label{fig:GHZ_Bloch_sphere}
\end{figure}

Since the symmetric subspace obtained by imposing the $S_3$ permutation symmetry on three qubits, has dimension two, we can map the three-qubit problem into an effective single-qubit representation. The geometric meaning of the resulting control protocol will become clear below. 

In order to ensure that the state stays in the symmetric subspace, we have to apply a symmetric rotation around the qubit $x$ axes:
\begin{equation}
U_F=e^{-i\frac{\theta}{2}(\sigma_{x,1} + \sigma_{x,2} + \sigma_{x,3})}.
\end{equation}
Then the symmetric subspace within which the dynamics takes place is spanned by
\begin{align}
|\tilde{0}\> &=\frac{1}{\sqrt{2}}(|000\>+|111\>)\\
|\tilde{1}\> &=\frac{1}{\sqrt{6}}(|100\>+|010\>+|001\>+|011\>+|101\>+|110\>)
\end{align}
The $|\tilde{0}\>$ basis is just the GHZ state we are trying to generate. Let us see how encoded operations are realized in this subspace. First, the rotation axis $\Sigma\equiv \sigma_{x,1} + \sigma_{x,2} + \sigma_{x,3}$ is equivalent to
\begin{align}
\tilde{\Sigma} &=
\begin{pmatrix}
\<\tilde{0}|\Sigma|\tilde{0}\> & \<\tilde{0} |\Sigma|\tilde{1}\>\\
\<\tilde{1}|\Sigma|\tilde{0}\> & \<\tilde{1}|\Sigma|\tilde{1}\>
\end{pmatrix}
=
\begin{pmatrix}
0 & \sqrt{3}\\
\sqrt{3} & 2
\end{pmatrix}\\ \nonumber
&=\tilde{I}+\sqrt{3}\tilde{X}-\tilde{Z}
\end{align}
where $\tilde{I}$, $\tilde{X}$, and $\tilde{Z}$ are the $2$ by $2$ identity, Pauli $X$ and $Z$ matrix. We can drop the identity here since it only gives a global phase. Then we have
\begin{equation}
\tilde{\Sigma}=\sqrt 3 \tilde{X} -\tilde{Z} = 2\hat{n}\cdot\vec{\tilde{\sigma}},
\end{equation}
where $\hat{n}=(\frac{\sqrt 3}{2}, 0, \frac{1}{2})$ is a unit vector lying in the $x-z$ plane which gives us the rotation axis, and $\vec{\tilde{\sigma}}$ is just the vector of the effective Pauli matrices. So in the effective single qubit picture, the rotation operator $\tilde{\sigma}$ and the corresponding rotation angle $\tilde{\theta}$ are given by
\begin{align} \label{eqn:optimalangle_Blochsphere}
\tilde{\sigma} &=\frac{\sqrt 3}{2}\tilde{X}-\frac{1}{2}\tilde{Z}\\
\tilde{\theta} &=2\theta
\end{align}
Note that the rotation angle in the effective single qubit space is equal to twice that in the original space. 

Now let us look at the measurement process, which is given by the encoded operator
\begin{equation}
\tilde{X}_{obs}\equiv
\begin{pmatrix}
\<\tilde{0}|X_{obs}|\tilde{0}\> & \<\tilde{0}|X_{obs}|\tilde{1} \>\\
\<\tilde{1}|X_{obs}|\tilde{0} \> & \<\tilde{1} |X_{obs}|\tilde{1} \>
\end{pmatrix}
=
\begin{pmatrix}
3 & 0\\
0 & -1
\end{pmatrix}
=\tilde{I}+2\tilde{Z}
\end{equation}
The evolution equation is given by Eq.~\eqref{eqn:measurement} of the main text. In our case, the Hamiltonian is zero, and for the terms deriving from the measurement, we have
\begin{align}
\mathcal{D}[\tilde{I} + 2\tilde{Z}] &=\mathcal{D}[2\tilde{Z}]\\
\mathcal{H}[\tilde{I}+2\tilde{Z}] &=\mathcal{H}[2\tilde{Z}].
\end{align}
Dropping the identity in the measured observable, we find that the measurement in the effective single qubit subspace is a measurement along the encoded $z$ axis, with a four-fold increase in the measurement strength, \textit{i.e.},
\begin{align}
\tilde{X}_{obs}=Z\\
\tilde{k} = 4k.
\end{align}
where $\tilde{k}$ is the effective measurement strength in the single qubit space.

It is easy to show that the initial state $|\psi\>_0=(\frac{1}{\sqrt{2}}(|0\>+|1\>))^{\otimes 3}$ becomes 
\begin{equation}
|\tilde{\psi}\>_0=\frac{1}{2}|\tilde{0} \>+\frac{\sqrt{3}}{2}|\tilde{1} \>
\end{equation} 
in the effective single qubit space, or   
in Bloch vector form,
\begin{equation}
\tilde{\rho}_0 =|\tilde{\psi}\>_0\<\tilde{\psi}|_0\sim (\frac{\sqrt 3}{2},0,-\frac{1}{2})
\end{equation}
The fidelity with respect to the GHZ state is then given in terms of this Bloch vector by
\begin{equation}
f\equiv \<\tilde{0}|\rho|\tilde{0}\>=\frac{1+\tilde{z}}{2},
\end{equation}
where $\tilde{z}$ is the $z$ component of the Bloch vector in the effective single qubit space.
Clearly the rotation angle has to be equal to either $0$ or $\frac{\pi}{2}$, in order to ensure that this fidelity is optimal at each time step.

\section{Tangle-based protocols}\label{app:non-markovian}
\begin{figure}    
    \includegraphics[width=0.5\textwidth]{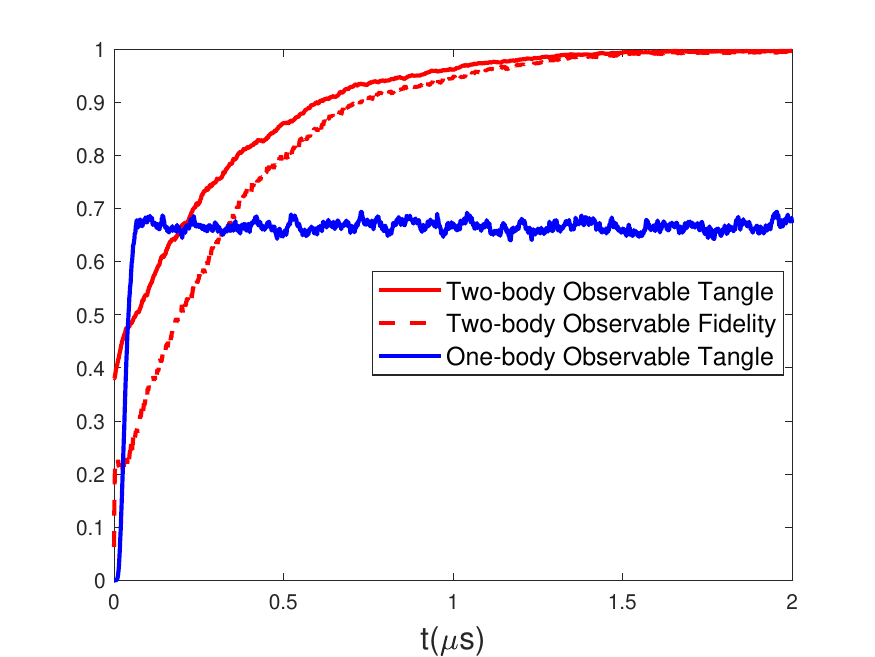}
     \caption{Time dependence of the three-tangle $\tau$ and the fidelity $\mathcal{F}_G$ for three qubits with the TEA protocol using feedback based on direct optimization of the three-tangle cost function $\tau$ (Eq.~\eqref{eq:tangle}).  The initial condition here is the complete superposition state, Eq.~\eqref{eq:fullsuperposition}. Red solid curve: three-tangle for state measured with two-body observable, averaged over $1000$ trajectories. Red dashed curve: fidelity with respect to GHZ state, same state evolution as the red solid curve. Blue solid curve : three-tangle of state measured with one-body observable, averaged over $100$ trajectories. For all trajectories the measurement strength was $k=1$ and time step $dt\ll \frac{1}{k}$.}
            \label{fig:Tangle_sym_a}
    \end{figure}
\begin{figure}
    \begin{subfigure}[t]{0.5\textwidth}            
            \includegraphics[width=\textwidth]{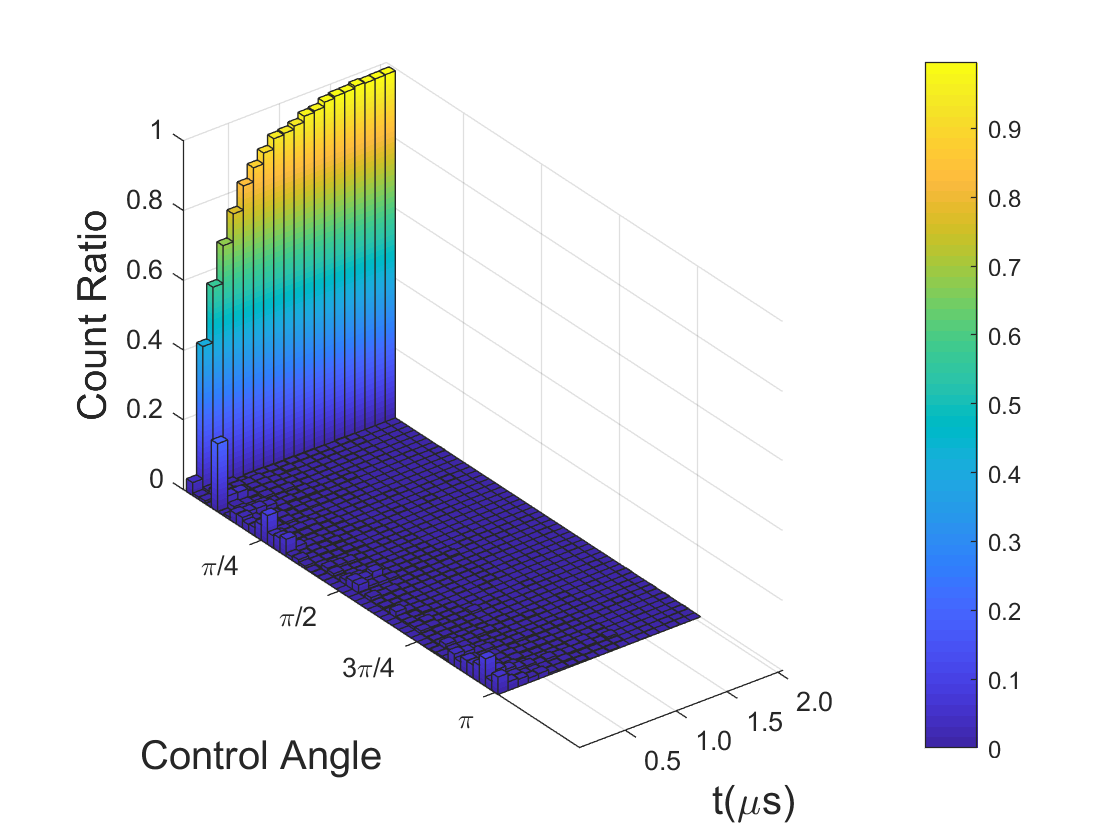}
            \caption{Histograms of control angle distribution as a function of time. $20$ snapshots of the angle distribution are taken between the initial and final times, with the control angle space $[0, \pi]$ divided into $50$ bins in each snapshot. The count ratio for the $0$ angle bin gradually increases to $1$, indicating approach to a steady state.}
            \label{fig:Tangle_sym_b}
    \end{subfigure}%

    \begin{subfigure}[t]{0.5\textwidth}
            \includegraphics[width=\textwidth]{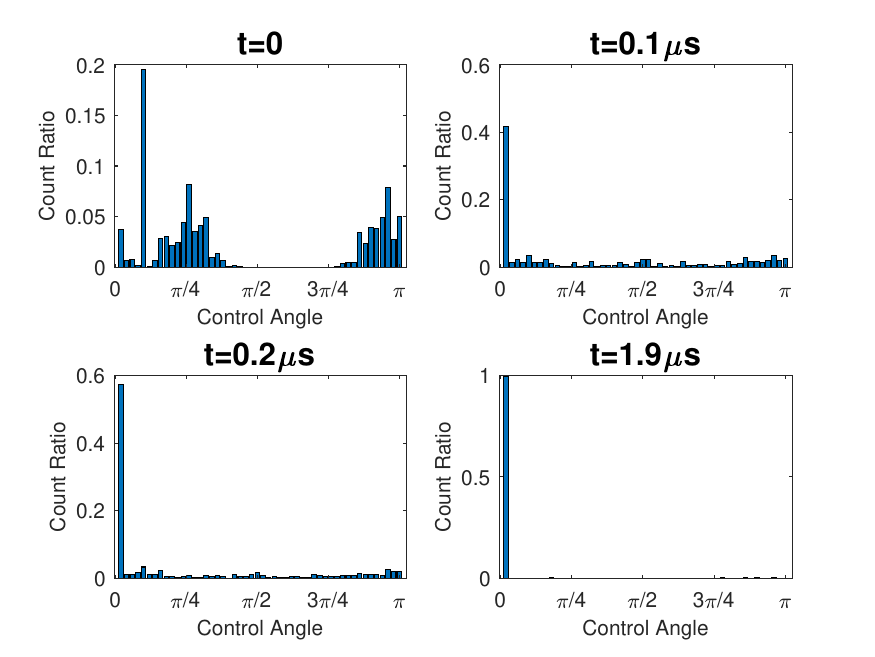}
            \caption{Four snapshots of the angular distribution, at times $t=0$, 0.1 $\mu$s, 0.2 $\mu$s, and 1.9 $\mu$s. The distribution rapidly becomes dominant in the first bin, and subsequent rotations of the state are very small.}
            \label{fig:Tangle_less_sym}
    \end{subfigure}
    \caption{Histograms of control angle distributions resulting from optimization of the three-tangle at different times. The total trajectory count is $1000$. The y-axis shows the normalized count in each bin.}
    \label{fig:Tangle}   
\end{figure}
Here we present an alternative locally optimal protocol for generation of a three-qubit GHZ state that is based on optimization of the three-tangle $\tau$, which provides a measure for tripartite entanglement~\cite{Coffman2000,Horodecki2009}. This protocol has the advantage that it not only differentiates between the two distinct types of tri-partite entanglement of three-qubit states, but is also invariant under local rotations of the state. Under this measure the GHZ state reaches the upper bound, with value $\tau = 1$, 
while all two-particle entanglements are zero~\cite{Sudbery2001,Acin2000}.
In the following we shall maximize the tangle under measurements of the symmetrized two-body operator $X^S_G$, Eq.~\eqref{eqn:observable_GHZ}.

We note first that starting from a pure state, the conditioned state after the weak measurement will still be pure. So if we avoid averaging the state along the evolution over measurement outcomes, the state will remain pure at all times.
This allows us to use the pure state definition of the three-tangle~\cite{Horodecki2009}
\begin{equation}\label{eq:tangle}
\tau = \tau(ABC) = \tau(A|BC) - \tau(A|B) - \tau(B|C),
\end{equation}
where the quantities on the right hand side (referred to as ``two-tangles") are given by squares of the relevant concurrences~\cite{Horodecki2009}. This considerably simplifies the determination of 
the feedback angles, since computing the tangle for a mixed state can be very difficult, involving determination of a convex roof extension~\cite{Cao2010}. 

Since the tangle is invariant under our feedback unitary operations, the maximization procedure used for the fidelity cost function in the main text of the paper does not work here. Instead, we determine the optimal angle by maximizing the expected increase in tangle 
\textit{after} measurement. The feedback angle at each infinitesimal time step is now computed as follows. 
At time $t$, the (pure) input state $|\psi\>_t$ is rotated using the feedback control operator
$U_F^G(t)$ (Eq.~\eqref{eqn:control_op_GHZ})
\begin{equation}
|\psi\>_t^c=U_F^G(t)|\psi\>_t,
\end{equation}
with the rotation angle parameter determined as described below.
We then make a weak measurement on the controlled state:
\begin{align}
|\psi\>_{t+dt}^c
&=\frac{\Omega_{dV}|\psi\>_t^c}{\lVert \Omega_{dV}|\psi\>_t^c \rVert} \\
&=\frac{\Omega_{dV} U_F^G(t)|\psi\>_t}{\lVert \Omega_{dV} U_F^G(t)|\psi\>_t \rVert}. \nonumber
\label{eq:evolvenonMarkov}
\end{align}
Note that we are now controlling the state {\it before} the measurement instead of {\it after} measurement: we choose to do this because a local rotation on a pure state will not change the value of the tangle for the state, so the tangle is not affected by the control.

Now the choice of rotation should not be determined by a particular measurement outcome that occurs {\it after} imposition of the control. Therefore in order to obtain the optimal rotation angle while avoiding issues of causality, 
we may simply average the tangle over all possible measurement outcomes and choose the control rotation as the value maximizing this average, i.e.,
\begin{align}
U_F^*(t)=\underset{U_F(t)}{argmax}\int dV\ \tau(|\psi\>_{t+dt}^c).
\end{align}
This requires sampling values of rotation angle and evaluating the average over measurement outcomes for each case.
The state is then evolved forward by acting with the measurement after the optimal rotation, yielding the evolution described by Eq.~\eqref{eq:evolvenonMarkov} with $U_F(t)$ replaced by $U_F^*(t)$.

Fig.~\ref{fig:Tangle_sym_a} shows that when this tangle-based protocol is implemented using the two-body measurement observable $X_{G}^{S}$, both the value of the three-tangle $\tau$ and the corresponding fidelity $\mathcal{F}_G$ appear to asymptotically reach a value of one, although on a slower timescale than the corresponding fidelity 
under the fidelity based approach (compare with Fig.~\ref{fig:GHZ_fidelity}). 
In contrast, when the tangle-based optimization is used with the non-symmetrized one-body observable $X_{G}$ for measurement, a significantly lower value of the tangle is obtained (not shown), with an asymptotic value of approximately 0.7 being reached. It is thus evident again that a protocol based on symmetrized two-body observable measurements significantly outperforms a protocol based on measurement with a non-fully symmetrized observable. 

The three-tangle $\tau$ is one of five non-trivial polynomial invariants that characterize normalized three-qubit states~\cite{Acin2000,Sudbery2001}. Our work suggests that optimization of multiple invariants might be useful for construction of feedback protocols to systematically generate arbitrarily entangled three-qubit states. For  three-qubit states, $\tau$ achieves its maximal value for the GHZ state and all other invariants automatically reach the boundary value. In this case, optimizing the tangle alone then guarantees that the other invariants reach the correct values for the state. This is not the case for other states in general. One alternative choice of cost function in more general situations is to use the sum of the squared differences between the invariants of the current state and the target state, which can act as a measure of the distance between the two states.

\end{appendix}

\bibliography{3qubitentangle}

\begin{thebibliography}{51}
\bibitem [{not()}]{note:measurem_probab}%
  \BibitemOpen
  \href@noop {} {}\bibinfo {note} {We note that for two qubits, entangled
  states can be produced with unit probability by measurements alone when full
  parity measurements are used. In this case, feedback provides additional
  speedup of the state generation~\cite{Hill2008,Martin2017a}.}\BibitemShut
  {Stop}%
\end{thebibliography}

\end{document}